\input{aipcheck}

\documentclass[
    ,final            
  ]
  {aipproc}

\layoutstyle{6x9}
\usepackage{color}

\begin{document}

\title{Dynamics and thermodynamics of systems with long-range interactions: interpretation of the different functionals}


\author{P.H. Chavanis}{
  address={ Laboratoire de Physique Th\'eorique, Universit\'e Paul
Sabatier, 118 route de Narbonne 31062 Toulouse, France}
}

\begin{abstract}
We discuss the dynamics and thermodynamics of systems with weak
long-range interactions. Generically, these systems experience a
violent collisionless relaxation in the Vlasov regime leading to a
(usually) non-Boltzmannian quasi stationary state (QSS), followed by a
slow collisional relaxation leading to the Boltzmann statistical
equilibrium state. These two regimes can be explained by a kinetic
theory, using an expansion of the BBGKY hierarchy in powers of $1/N$,
where $N$ is the number of particles. We discuss the physical meaning
of the different functionals appearing in the analysis: the Boltzmann
entropy, the Lynden-Bell entropy, the ``generalized'' entropies
arising in the reduced space of coarse-grained distribution functions,
the Tsallis entropy, the generalized $H$-functions increasing during
violent relaxation (not necessarily monotonically) and the convex
Casimir functionals used to settle the formal nonlinear dynamical
stability of steady states of the Vlasov equation. We show the
connection between the different variational problems associated with
these functionals. We also introduce a general class of nonlinear mean
field Fokker-Planck (NFP) equations that can be used as numerical
algorithms to solve these constrained optimization problems.
\end{abstract}

\maketitle


\section{1. Introduction}

Recently there was a renewed interest for the dynamics and
thermodynamics of systems with long-range interactions
\cite{dauxois}. These systems are numerous in Nature and appear in
different domains of physics, astrophysics, fluid mechanics and
biology. Systems with long-range interactions have very particular
properties that contrast from those of more familiar systems with
short-range interactions like gases and neutral plasmas. First of all,
they can organize spontaneously into ``coherent structures'' in
physical space. This corresponds to stars, galaxies, clusters of
galaxies...  in astrophysics, jets and vortices (like the gulf stream
or jovian vortices) in two-dimensional geophysical turbulence, clumps
and filaments (interpreted as the initiation of a vasculature) in the
chemotaxis of biological populations, clusters in the HMF
model... These systems have a very special thermodynamic limit
corresponding to $N\rightarrow +\infty$ in such a way that the
coupling constant tends to zero ($k\sim 1/N\rightarrow 0$) while the
volume remains unity ($V\sim 1$). This limit describes systems with
{\it weak} long-range interactions. For these systems, the mean field
approximation becomes exact for $N\rightarrow +\infty$.  Generically,
these systems display two successive types of relaxation: a violent
collisionless relaxation leading to a Quasi Stationary State (QSS) and
a slow collisional relaxation tending to the Boltzmann distribution of
statistical equilibrium.  The QSS is a stationary solution of the
Vlasov equation on the coarse-grained scale. It is usually described
by a non-Boltzmannian distribution and its lifetime diverges with
$N$. Mathematically, the Bolzmann equilibrium is obtained when the
$t\rightarrow +\infty$ limit is taken before the $N\rightarrow
+\infty$ limit and the QSS is obtained when the $N\rightarrow +\infty$
limit is taken before the $t\rightarrow +\infty$ limit. These two
regimes can be explained by a kinetic theory, using an expansion of
the BBGKY hierarchy in powers of $1/N$ with $N\rightarrow +\infty$
\cite{paper1,paper2,paper3,paper4}. Other peculiar behaviors in the
dynamics and thermodynamics of systems with long-range interactions
have been evidenced: negative specific heats in the microcanonical
ensemble \cite{lbw,thirring,gross,kiessling}, ensemble inequivalence
\cite{paddy,ellis,aa3,touchette}, numerous types of phase transitions
persisting at the thermodynamic limit \cite{bb,ijmpb}, non-trivial
dependence of the collisional relaxation time with the number $N$ of particles
\cite{yamaguchi,campa}, algebraic decay of the correlation functions
\cite{corr,rp,bd,kinvortex}, front structure and slow relaxation of 
the velocity tails
\cite{cl,kinvortex}, metastable states whose lifetimes diverge exponentially
with $N$ \cite{rm,lifetime}, out-of-equilibrium phase transitions
\cite{JFM,epjb,anto,antoPRL,vb}, kinetic blocking due 
to the absence of resonances
\cite{kinvortex,bbgky}, curious effects due to spatial inhomogeneity \cite{curious}, non-ergodic behaviors \cite{latora,prt}, glassy dynamics \cite{glass}
etc... Certainly, the physical richness of these systems (Pandora's
box) will lead to further investigations and applications.

In this paper, we discuss the meaning of the different functionals
appearing in the study of the dynamics and thermodynamics of systems
with weak long-range interactions. The Boltzmann entropy describes the
statistical equilibrium state resulting from a collisional
relaxation. The Lynden-Bell entropy \cite{lb} and the ``generalized''
entropic functionals arising in the reduced space of coarse-grained
distributions \cite{super} describe the QSS resulting from a complete
violent collisionless relaxation. The maximization of these entropic
functionals rely on an assumption of ergodicity (efficient mixing).
Tsallis entropies
\cite{tsallis}  have been proposed as an alternative when the system
has a non-ergodic behaviour so that violent relaxation is
incomplete. Incomplete relaxation may also be understood by developing
a kinetic theory of the process of violent relaxation
\cite{csr,quasivlasov,paper4}. Generalized $H$-functions \cite{thlb} are functionals of
the coarse-grained distribution function that increase during violent
relaxation (not necessarily monotonically). Finally, convex Casimir
functionals are used to settle the formal nonlinear dynamical
stability of steady states of the Vlasov equation
\cite{ellisgeo,aaantonov}.  Although the above-mentioned functionals may
have similar mathematical forms, they have very different physical
interpretations.

\section{2. Systems with weak long-range interactions} 

\subsection{2.1. The $N$-body Hamiltonian system} 

We consider an isolated system of particles in interaction
whose dynamics is fully described by the Hamilton equations
\begin{eqnarray}
\label{eqb1} m{d{\bf r}_{i}\over dt}={\partial H\over\partial {\bf v}_{i}}, \qquad m{d{\bf v}_{i}\over dt}=-{\partial H\over\partial {\bf r}_{i}}, \nonumber\\
H={1\over 2}\sum_{i=1}^{N}m{v_{i}^{2}}+m^{2}\sum_{i<j}u({\bf r}_{i}-{\bf r}_{j}),
\end{eqnarray}
where $u_{ij}=u({\bf r}_{i}-{\bf r}_{j})$ is a binary potential of
interaction depending only on the absolute distance $|{\bf r}_{i}-{\bf
r}_{j}|$ between the particles. Basically, the evolution of the $N$-body distribution function $P_N({\bf
r}_{1},{\bf v}_{1},...,{\bf r}_{N},{\bf v}_{N},t)$ is
governed by the Liouville equation
\begin{equation}
\label{eqb2} {\partial P_{N}\over\partial t}+\sum_{i=1}^{N}\biggl
({\bf v}_{i}\cdot {\partial P_{N}\over\partial {\bf r}_{i}}+{\bf
F}_{i}\cdot {\partial P_{N}\over\partial {\bf v}_{i}}\biggr
)=0,
\end{equation}
where
\begin{equation}
\label{eqb3}
{\bf F}_{i}=-{\partial \Phi\over\partial {\bf r}_{i}}=-m\sum_{j\neq i}{\partial u_{ij}\over\partial {\bf r}_{i}}=\sum_{j\neq i}{\bf F}(j\rightarrow i),
\end{equation}
is the force by unit of mass experienced by particle $i$ due to the
interaction with the other particles. The Liouville equation is
equivalent to the $N$-body Hamiltonian system (\ref{eqb1}).  Now, the
basic postulate of statistical mechanics states that: {\it At
statistical equilibrium, all the microscopic configurations that are
accessible (i.e. that have the correct value of the energy) are
equiprobable}. There is no guarantee that the dynamics will lead the
system to that ``uniform'' state. This relies upon a hypothesis of
ergodicity which may not always be fulfilled for complex systems (some
regions of the $\Gamma$-phase space could be more probable than
others). This is why it is important to develop a kinetic theory in
order to vindicate (or not!) this result. In the sequel, following the
wisdom of ordinary statistical mechanics, we shall assume that all the
accessible microscopic configurations are equiprobable at equilibrium
(for $t\rightarrow +\infty$) but we must keep in mind the importance
of this postulate.

At statistical  equilibrium, the  $N$-body distribution is given by the
microcanonical distribution expressing the equiprobability of the accessible configurations
\begin{equation}
\label{eqb4} P_{N}({\bf r}_{1},{\bf v}_{1},...,{\bf r}_{N},{\bf v}_{N})={1\over
g(E)} \delta(E-H({\bf r}_{1},{\bf v}_{1},...,{\bf r}_{N},{\bf v}_{N})).
\end{equation}
Since $\int P_{N}\prod_{i}d{\bf
r}_{i}d{\bf v}_{i}=1$, the density of states with energy $E$ is given
by
\begin{equation}
\label{eqb5} g(E)=\int  \delta(E-H({\bf r}_{1},{\bf v}_{1},...,{\bf r}_{N},{\bf v}_{N}))\prod_{i}d{\bf
r}_{i}d{\bf v}_{i}.
\end{equation}
This is the normalization factor of the $N$-body distribution function
(\ref{eqb4}). The microcanonical entropy of the system is defined by
$S(E)=\ln g(E)$ and the microcanonical temperature by
$\beta(E)=1/T(E)=\partial S/\partial E$ (we take the Boltzmann
constant $k_{B}=1$).  We introduce the reduced probability
distributions
\begin{equation}
\label{eqb6}
P_{j}({\bf x}_{1},...,{\bf x}_{j})=\int P_{N}({\bf x}_{1},...,{\bf x}_{N})d{\bf x}_{j+1}...d{\bf x}_{N},
\end{equation}
where ${\bf x}=({\bf r},{\bf v})$. For identical particles, the
average density in $\mu$-space is related to the one-body
distribution function by
\begin{equation}
\label{eqb7} f({\bf r},{\bf v})=\langle \sum_{i=1}^{N}m\delta({\bf r}-{\bf r}_{i})\delta({\bf v}-{\bf v}_{i})\rangle =Nm P_{1}({\bf r},{\bf v}),
\end{equation}
and the average value of the energy is
\begin{eqnarray}
\label{eqb8}
E=\langle H\rangle=Nm\int P_{1}({\bf r},{\bf v}){v^{2}\over 2}d{\bf r}d{\bf v}
\nonumber\\
+{1\over 2}N(N-1)m^{2}\int u({\bf r}-{\bf r}')P_{2}({\bf r},{\bf v},{\bf r}',{\bf v}')d{\bf r}d{\bf v}d{\bf r}'d{\bf v}'.
\end{eqnarray}
By differentiating the defining relation for $P_j$ and using
Eq. (\ref{eqb4}), we can obtain an equilibrium BBGKY-like hierarchy
for the reduced distributions \cite{paper1}:
\begin{equation}
\label{eqb9}
{\partial P_{j}\over\partial {\bf r}_{1}}=-{1\over g(E)}{\partial\over\partial E}\biggl\lbrack g(E)P_{j}\biggr\rbrack\sum_{i=2}^{j}m^{2}{\partial u_{1,i}\over\partial {\bf r}_{1}}-(N-j)m^{2}\int
{\partial u_{1,j+1}\over\partial {\bf r}_{1}}{1\over g(E)}{\partial\over\partial E}\biggl\lbrack g(E)P_{j+1}\biggr\rbrack   d{\bf x}_{j+1},
\end{equation}
\begin{equation}
\label{eqb10}
{\partial P_{j}\over\partial {\bf v}_{1}}=-{1\over g(E)}{\partial\over\partial E}\biggl\lbrack g(E)P_{j}\biggr\rbrack m{\bf v}_{1}.
\end{equation}
This is the counterpart of the equilibrium hierarchy in plasma physics. It
is however more complex in the present situation because it has been
derived in the {\it microcanonical} ensemble. Since statistical ensembles
are generically inequivalent for systems with long-range
interactions, we must formulate the problem in the microcanonical
ensemble which is the fundamental one. We note that the terms involving the derivative of the density of
states can be split in two parts according to
\begin{equation}
\label{eqb11}
{1\over g(E)}{\partial\over\partial E}\biggl\lbrack g(E)P_{j}\biggr\rbrack=\beta P_{j}+{\partial P_{j}\over\partial E}.
\end{equation}
The terms with the $E$ derivative would not have emerged if we had
started from the Gibbs canonical distribution \cite{paper1}.

\subsection{2.2. Thermodynamic limit and mean-field approximation}

We define the thermodynamic limit as $N\rightarrow +\infty$ in such a
way that the normalized energy $\epsilon=E/(u_{*}N^{2}m^{2})$ and the
normalized temperature $\eta=\beta Nm^{2} u_{*}$ are fixed, where
$u_{*}$ represents the typical value of the potential of interaction.
Since the normalized coupling constant $\beta m^{2}u_{*}=\eta/N\sim
1/N$ goes to zero for $N\rightarrow+\infty$, we are studying systems
with {\it weak} long-range interactions. For 3D self-gravitating
systems ($u=-G/|{\bf r}-{\bf r}'|$, $u_{*}=G/R$, $\eta={\beta GMm/
R}$, $\epsilon={ER/ GM^{2}}$), for 2D point vortices
\footnote{Two-dimensional point vortices (with circulation $\gamma$)
are special because they have no inertia. Their kinetic theory, which
must take into account this particularity, is developed in
\cite{bbgky}.}  ($u=-{1\over 2\pi}\ln |{\bf r}-{\bf r}'|$, $u_{*}=1$,
$\eta={\beta N\gamma^{2}}$, $\epsilon={E/\Gamma^{2}}$) and for the HMF
model ($u=-{k\over 2\pi}\cos(\theta-\theta')$, $u_{*}=k$, $\eta={\beta
kN}$, $\epsilon={E/kN^{2}}$) we recover the dimensionless parameters
already introduced in the literature. In general, the potential of
interaction is written as $u({\bf r}_{ij})=k\tilde{u}({\bf r}_{ij})$
where $k$ is the coupling constant (e.g., $G$ for self-gravitating
systems or $k$ for the HMF model).  The dynamical time is $t_{D}\sim
R/v_{typ}\sim R/\sqrt{N m u_*}$ where the typical velocity
$v_{typ}\sim (Nmu_{*})^{1/2}$ has been obtained by equating the
kinetic energy $\sim N m v^{2}$ and the potential energy $\sim N^2
m^{2} u_*$. For a long-range interaction, we have $u_*=k/R^{\alpha}$
where $\alpha$ is less than the dimension $d$ of the system. Then, we
get $t_{D}\sim R^{(2+\alpha-d)/2}/\sqrt{k\rho}$ where $\rho\sim M/V$
is the average density.  By a suitable normalization of the
parameters, the proper thermodynamic limit for systems with long-range
interactions is such that the coupling constant behaves like $k\sim
u_{*}\sim 1/N$, while $m\sim 1$, $\beta\sim 1$, $E/N\sim 1$ and $V\sim
1$. This implies that $|{\bf r}|\sim 1$, $|{\bf v}|\sim 1$, $u\sim
1/N$, $|{\bf F}(j\rightarrow i)|\sim 1/N$ and $t_{D}\sim 1$. For
example, in the case of self-gravitating systems, this scaling
corresponds to $G\sim 1/N$ while $m\sim 1$, $\beta\sim 1$, $E/N\sim
1$, $V\sim 1$, $t_{D}\sim 1/\sqrt{G\rho}\sim 1$ and in the case of the
HMF model, it corresponds to $k\sim 1/N$ while $\beta\sim 1$, $E/N\sim
1$, $V\sim 1$, $t_{D}\sim 1/\sqrt{k\rho}\sim 1$ (for self-gravitating
systems, we could also consider the dilute limit $N\rightarrow
+\infty$ with $N/R\sim 1$ while $G\sim 1$, $m\sim 1$, $\beta\sim 1$,
$E/N\sim 1$ \cite{sanchez} but in that case we notice that $t_{D}\sim
N$). Using another normalization of the parameters, the thermodynamic
limit is such that the mass of the particles behaves like $m\sim 1/N$
while $k\sim u_{*}\sim 1$, $\beta\sim N$, $E\sim 1$, $V\sim 1$ and
$t_{D}\sim 1$. In this scaling, the total mass $M\sim Nm$ is of order
unity. For example, in the case of self-gravitating systems, this
scaling corresponds to $m\sim 1/N$ while $G\sim 1$, $\beta\sim N$,
$E\sim 1$, $V\sim 1$, $t_{D}\sim 1/\sqrt{G\rho}\sim 1$ and in the case
of 2D vortices to $\gamma\sim 1/N$, while $\beta\sim N$, $E\sim 1$,
$V\sim 1$, $t_{D}\sim 1/\omega\sim R^{2}/\Gamma\sim 1$ (where $\Gamma=N\gamma$ is the total circulation).

The first two equations of the equilibrium BBGKY-like hierarchy are 
\begin{eqnarray}
\label{tl1} {\partial P_{1}\over\partial {\bf r}_{1}}({\bf x}_{1})=-(N-1) m^{2}\int {\partial u_{12}\over\partial {\bf r}_{1}} \left (\beta P_{2}+\frac{\partial P_{2}}{\partial E}\right )d{\bf x}_{2},
\end{eqnarray}
\begin{eqnarray}
\label{tl2}
{\partial P_{2}\over\partial {\bf r}_{1}}({\bf x}_{1},{\bf
x}_{2})=- m^{2}\left (\beta P_{2}+\frac{\partial P_{2}}{\partial E}\right ) {\partial
u_{12}\over\partial {\bf r}_{1}}- (N-2) m^{2}\int {\partial
u_{13}\over\partial {\bf r}_{1}} \biggl (\beta P_{3}+{\partial P_{3}\over \partial E}\biggr ) d{\bf x}_{3}.
\end{eqnarray}
We note that the ratio of $\partial
P_{j}/\partial E$ on $\beta P_{j}$ is of order
$1/(E\beta)=1/(\epsilon\eta N)$. Therefore, in the thermodynamic limit
$N\rightarrow +\infty$ with $\epsilon$, $\eta$ fixed, the second term
in Eq. (\ref{eqb11}) is always negligible with respect to the first.
We now decompose the two- and three-body distribution functions in the
suggestive form
\begin{equation}
\label{tl3}
P_{2}({\bf x}_{1},{\bf x}_{2})=P_{1}({\bf x}_{1})P_{1}({\bf x}_{2})+P_{2}'({\bf x}_{1},{\bf x}_{2}),
\end{equation}
\begin{eqnarray}
\label{tl4}
P_{3}({\bf x}_{1},{\bf x}_{2},{\bf x}_{3})=P_{1}({\bf x}_{1})P_{1}({\bf x}_{2})P_{1}({\bf x}_{3})+P_{2}'({\bf x}_{1},{\bf x}_{2})P_{1}({\bf x}_{3})\nonumber\\
+P_{2}'({\bf x}_{1},{\bf x}_{3})P_{1}({\bf x}_{2})+P_{2}'({\bf x}_{2},{\bf x}_{3})P_{1}({\bf x}_{1})+P_{3}'({\bf x}_{1},{\bf x}_{2},{\bf x}_{3}).
\end{eqnarray}
This decomposition corresponds to the first terms of the Mayer
expansion in plasma physics.  The non-trivial correlations $P_{j}'$
are called the cumulants. We can now substitute these decompositions
in the equilibrium BBGKY-like hierarchy. The first equation
(\ref{tl1}) of the hierarchy can be written
\begin{eqnarray}
\label{tl5} {\partial P_{1}\over\partial {\bf r}_{1}}({\bf x}_{1})=-\beta (N-1) m^{2} P_{1}({\bf x}_{1})\int P_{1}({\bf x}_{2}){\partial u_{12}\over\partial {\bf r}_{1}}d{\bf x}_{2}\nonumber\\
-\beta (N-1) m^{2}\int P_{2}'({\bf x}_{1},{\bf x}_{2}){\partial u_{12}\over\partial {\bf r}_{1}}d{\bf x}_{2}-(N-1) m^{2}\int {\partial u_{12}\over\partial {\bf r}_{1}}{\partial P_{2}\over\partial E}({\bf x}_{1},{\bf x}_{2})d{\bf x}_{2}.
\end{eqnarray}
In the thermodynamic limit defined previously, it can be shown that
the cumulants $P_{n}'$ are of order $N^{-(n-1)}$ \cite{paper1}. In
particular, we have
\begin{equation}
\label{tl7}
P_{2}({\bf x}_{1},{\bf x}_{2})=P_{1}({\bf x}_{1})P_{1}({\bf x}_{2})+O(1/N).
\end{equation}
Therefore, the mean field approximation is exact for $N\rightarrow
+\infty$. In the thermodynamic limit, the two-body correlation
function is the product of two one-body distribution functions:
$P_{2}({\bf x}_{1},{\bf x}_{2})\simeq P_{1}({\bf x}_{1})P_{1}({\bf
x}_{2})$.

\subsection{2.3. The mean field equilibrium distribution}

Taking the limit $N\rightarrow +\infty$ and using Eq. (\ref{tl7}), the first equation (\ref{tl5}) of the equilibrium BBGKY-like hierarchy becomes
\begin{equation}
\label{mfeq1}
\nabla\rho({\bf r})=-\beta m\rho({\bf r})\nabla\int\rho({\bf r}')u({\bf r}-{\bf r}')d{\bf r}',
\end{equation}
where $\rho({\bf r})=N m P_{1}({\bf r})$ is the spatial density. After integration, this can be written in the form of a meanfield  Boltzmann distribution
\begin{equation}
\label{mfeq2} \rho({\bf r})=A' e^{-\beta m\Phi({\bf r})},
\end{equation}
where
\begin{equation}
\label{mfeq3} \Phi({\bf r})=\int \rho({\bf r}')u({\bf r}-{\bf r}')d{\bf r}',
\end{equation}
is the potential produced self-consistently by the smooth distribution
of particles. Therefore, the equilibrium density profile of the
particles is determined by an {\it integrodifferential} equation
(\ref{mfeq1}) or (\ref{mfeq2})-(\ref{mfeq3}). On the other hand, for
$N\rightarrow +\infty$, Eq. (\ref{eqb10}) leads to
\begin{equation}
\label{mfeq4}{\partial P_{1}\over\partial {\bf v}_{1}}=-\beta m P_{1}{\bf v}_{1},
\end{equation}
so that the velocity distribution $\phi({\bf v})=NmP_{1}({\bf v})$ is maxwellian
\begin{equation}
\label{mfeq5}\phi({\bf v})=A'' e^{-\beta m {v^{2}\over 2}}.
\end{equation}
Combining Eqs. (\ref{mfeq2}) and (\ref{mfeq5}) and introducing the
distribution function $f({\bf r},{\bf v})=NmP_{1}({\bf r},{\bf
v})$, we get the mean-field
Maxwell-Boltzmann distribution function
\begin{equation}
\label{mfeq6} f({\bf r},{\bf v})=Ae^{-\beta m({v^{2}\over 2}+\Phi({\bf r}))},
\end{equation}
where $\Phi({\bf r})$ is given by Eq. (\ref{mfeq3}) and $\rho=\int f d{\bf v}$.
If we define the entropy by
\begin{equation}
\label{eea1} S_{N}=-\int P_{N}\ln P_{N}d{\bf x}_{1}...d{\bf x}_{N},
\end{equation}
and use the mean-field approximation
\begin{equation}
\label{eea2} P_{N}({\bf x}_{1},...,{\bf x}_{N})=P_{1}({\bf x}_{1})...P_{1}({\bf x}_{N}),
\end{equation}
valid at the thermodynamic limit $N\rightarrow +\infty$ defined previously, we obtain
\begin{equation}
\label{eea3} S=-N\int P_{1}({\bf x})\ln P_{1}({\bf x})d{\bf x}.
\end{equation}
In terms of the distribution function $f=NmP_{1}$, we get
\begin{equation}
\label{eea4} S=-\int {f\over m}\ln\biggl ({f\over Nm}\biggr )d{\bf r}d{\bf v}.
\end{equation}
On the other hand, in the mean-field approximation, the average energy
(\ref{eqb8}) is given by
\begin{equation}
\label{eea6} E=Nm\int P_{1}{v^{2}\over 2}d{\bf r}d{\bf v}+{1\over 2}N^{2}m^{2}\int P_{1}({\bf r})u({\bf r}-{\bf r}')P_{1}({\bf r}')d{\bf r}d{\bf r}'.
\end{equation}
This can also be written
\begin{equation}
\label{eea7} E=\int f {v^{2}\over 2}d{\bf r}d{\bf v}+{1\over 2}\int  \rho({\bf r})\Phi({\bf r})d{\bf r}.
\end{equation}
The mass is given by 
\begin{equation}
\label{eea8} M=\int f d{\bf r} d{\bf v}.
\end{equation}
Using these relations, we can relate the inverse temperature $\beta$
and the constant $A$ in the mean field Maxwell-Boltzmann distribution
(\ref{mfeq6}) to the energy $E$ and the mass $M$.

\subsection{2.4. Statistical equilibrium state: the Boltzmann entropy}

We wish to determine the {\it most probable} distribution of particles at
statistical equilibrium by using a combinatorial analysis, assuming
that all accessible microstates (with given energy $E$ and
mass $M$) are equiprobable. To that purpose, we divide the
$\mu$-space $\lbrace {\bf r},{\bf v}\rbrace$ into a very large number
of microcells with size $h$. We do not put any exclusion, so that a
microcell can be occupied by an arbitrary number of particles.  We
shall now group these microcells into macrocells each of which
contains many microcells but remains nevertheless small compared to
the phase-space extension of the whole system. We call $\nu$ the
number of microcells in a macrocell. Consider the configuration
$\lbrace n_{i} \rbrace$ where $n_{i}$ is the number of particles 
in the macrocell $i$. Using the standard combinatorial
procedure introduced by Boltzmann, the probability of the state
$\lbrace n_{i}\rbrace$, i.e. the number of microstates corresponding
to the macrostate $\lbrace n_{i}\rbrace$, is given by
\begin{equation}
\label{bol1} W(\lbrace n_{i}\rbrace)=\prod_{i}N!{\nu^{n_{i}}\over n_{i}!}.
\end{equation}
This is the Maxwell-Boltzmann statistics. As is customary, we define
the entropy of the state $\lbrace n_{i} \rbrace$ by
\begin{equation}
S(\lbrace n_{i} \rbrace)=\ln W(\lbrace n_{i} \rbrace). \label{bol2}
\end{equation}
It is convenient here to return to a representation in terms of the
distribution function giving the phase-space density in the $i$-th
macrocell: $f_{i}=f({\bf r}_i,{\bf v}_i)={n_{i}m/\nu h^{d}}$ (where
$d$ is the dimension of space). Using the Stirling formula $\ln
n!\simeq n\ln n-n$ for $n\gg 1$, we have
\begin{equation}
\ln W(\lbrace n_{i} \rbrace)=-\sum_{i}n_{i}\ln {n_{i}}=-\sum_{i}\nu h^{d}{f_{i}\over m}\ln {f_{i}\over m}. \label{bol3}
\end{equation}
Passing to the continuum limit
$\nu\rightarrow 0$, we obtain the usual expression of the Boltzmann
entropy 
\begin{equation}
\label{bol4} S_{B}[f]=-\int {f\over m}\ln {f\over m} d{\bf r}d{\bf v},
\end{equation}
up to some unimportant additive constant.  Assuming ergodicity, the
statistical equilibrium state, corresponding to the most probable
distribution of particles (i.e. the macrostate that is the most
represented at the microscopic level), is obtained by maximizing the
Boltzmann entropy (\ref{bol4}) while conserving the total mass
(\ref{eea8}) and the total energy (\ref{eea7}). Introducing Lagrange
multipliers and writing the variational principle in the form
\begin{equation}
\label{bol8} \delta S_{B}-\beta\delta E-\alpha\delta M=0,
\end{equation}
we obtain the meanfield Maxwell-Boltzmann distribution
\begin{equation}
\label{bol9} f=A e^{-\beta m ({v^{2}\over 2}+\Phi)}.
\end{equation}
This returns the equations obtained in Sec. 2.3. The potential $\Phi$
is obtained by solving the integrodifferential equation
(\ref{mfeq2})-(\ref{mfeq3}) and the Lagrange multipliers $A$ and
$\beta$ must be related to the constraints $M$ and $E$ using
Eqs. (\ref{eea8})-(\ref{eea7}). Then, we have to make sure that the
distribution is a {\it maximum} of $S_{B}$ at fixed mass and energy,
not a minimum or a saddle point. Therefore, this method gives a
condition of {\it thermodynamical stability} which was not obtained in
Sec. 2.3 (to get a similar relation, we need to consider the next
order term in the expansion in $1/N$ of the equilibrium BBGKY-like
hierarchy).

The maximization problem 
\begin{eqnarray}
\label{bol10}
\max_{f}\quad \lbrace S_{B}[{f}]\quad |\quad E[{f}]=E, \ M[{f}]=M \rbrace,
\end{eqnarray}
corresponds to a condition of microcanonical stability. Alternatively, the minimization problem
\begin{eqnarray}
\label{bol11}
\min_{f}\quad \lbrace F_{B}[{f}]=E[f]-TS_{B}[f] \quad |\quad \ M[{f}]=M \rbrace,
\end{eqnarray}
corresponds to a condition of canonical stability, where $F_{B}[f]$ is
the Boltzmann free energy. The optimization problems (\ref{bol10}) and
(\ref{bol11}) have the same critical points (cancelling the first
order variations of the thermodynamical potential). On the other hand,
canonical stability (\ref{bol11}) implies microcanonical stability
(\ref{bol10}): if $f({\bf r},{\bf v})$ is a minimum of $F_{B}[f]$ at
fixed mass, then it is a maximum of $S_{B}[f]$ at fixed mass and
energy. However, the converse is wrong in case of {\it ensemble
inequivalence}. This means that the ensemble of solutions of
(\ref{bol11}) is included in the ensemble of solutions of
(\ref{bol10}) but the two ensembles may not coincide for systems with
long-range interactions \footnote{Since energy is {\it non-additive}
for systems with long-range interactions, the canonical ensemble
cannot be used to study a subpart of the system
\cite{paddy,dauxois}. Therefore, the problem (\ref{bol11}) has no physical
justification for an {\it isolated} Hamiltonian system interacting via
purely long-range forces. However, the canonical ensemble is justified
for {\it dissipative} systems, such as Brownian systems with
long-range interactions, that experience a friction force and a
stochastic force (modelling short-range interactions) in addition to
long-range forces \cite{paper1}. Here, we shall only consider isolated
Hamiltonian systems so that the fundamental ensemble is the
microcanonical ensemble. However, the minimization problem
(\ref{bol11}) can be useful mathematically since it offers a {\it
sufficient} condition of microcanonical stability (since (\ref{bol11})
$\Rightarrow$ (\ref{bol10})) as discussed in the text.}. 
Ensemble inequivalence was first encountered in astrophysics
\cite{lbw,thirring} where it was realized that configurations with
negative specific heats are allowed in the microcanonical ensemble but
not in the canonical ensemble (see reviews in
\cite{paddy,ijmpb}). This notion of ensemble inequivalence has been
developed and formalized recently at a general level
\cite{ellis,touchette,bb}. The condition of ensemble equivalence 
or ensemble inequivalence, related to the concavity of the entropy
$S(E)$, is discussed in detail in these papers. Ensemble
equivalence/inequivalence can also be deduced from the study of the
series of equilibria $\beta(E)$ using the Poincar\'e criterion
\cite{katz,ijmpb}.

\section{3. Kinetic theory from the BBGKY hierarchy}

In this section, we derive a general kinetic equation (\ref{general})
for Hamiltonian systems with weak long-range interactions. We start
from the BBGKY hierarchy and use a systematic expansion of the
solutions of the equations of this hierarchy in powers of $1/N$ in a
proper thermodynamic limit $N\rightarrow +\infty$. The kinetic
equation (\ref{general}) is valid at order $O(1/N)$.

\subsection{3.1. The $1/N$ expansion}

From the Liouville equation (\ref{eqb2}) we can construct the complete BBGKY hierarchy for the reduced distribution functions (\ref{eqb6}). It reads
\begin{equation}
\label{h1} {\partial P_{j}\over\partial t}+\sum_{i=1}^{j}{\bf v}_{i}{\partial
P_{j}\over\partial {\bf r}_{i}}+\sum_{i=1}^{j}\sum_{k=1,k\neq i}^{j} {\bf F}(k\rightarrow i){\partial P_{j}\over \partial {\bf v}_{i}}+(N-j)\sum_{i=1}^{j}\int d{\bf x}_{j+1}{\bf F}(j+1\rightarrow i){\partial P_{j+1}\over\partial {\bf v}_{i}}=0.
\end{equation}
This hierarchy of equations is not closed since the equation for the
one-body distribution $P_{1}({\bf x}_{1},t)$ involves the two-body
distribution $P_{2}({\bf x}_{1},{\bf x}_{2},t)$, the equation for the
two-body distribution $P_{2}({\bf x}_{1},{\bf x}_{2},t)$ involves the
three-body distribution $P_{3}({\bf x}_{1},{\bf x}_{2},{\bf
x}_{3},t)$, and so on... The idea is to close the hierarchy by using a
systematic expansion of the solutions in powers of $1/N$ in the
thermodynamic limit $N\rightarrow +\infty$. Considering the scaling of
the terms in each equation of the hierarchy, we argue that there
exists solutions of the whole BBGKY hierarchy such that the
correlation functions $P_{j}'$ scale like $1/N^{j-1}$ at any time
\cite{paper3}. This implicitly assumes that the initial condition has
no correlation, or that the initial correlations respect this scaling
(if there are strong correlations in the initial state, like
``binaries'', the kinetic theory will be different from the one
developed in the sequel).  If this scaling is satisfied, we can
consider an expansion of the solutions of the equations of the
hierarchy in terms of the small parameter $1/N$. This is similar to
the expansion in terms of the plasma parameter made in plasma
physics. However, in plasma physics the systems are spatially
homogeneous while, in the present case, we shall take into account
spatial inhomogeneity. This brings additional terms in the kinetic
equations that are absent in plasma physics. Therefore, strictly
speaking, the hierarchy that we consider is different from the
ordinary BBGKY hierarchy.  If we introduce the notations $f=NmP_{1}$
(distribution function) and $g=N^{2}P_{2}'$ (two-body correlation
function), we get at order $1/N$ \cite{paper3}:
\begin{eqnarray}
{\partial f_{1}\over\partial t}+{\bf v}_{1}{\partial f_{1}\over\partial {\bf r}_{1}}+\frac{N-1}{N}\langle {\bf F}\rangle_{1} {\partial f_{1}\over \partial {\bf v}_{1}}=-m {\partial \over\partial {\bf v}_{1}}\int {\bf F}(2\rightarrow 1)g({\bf x}_{1},{\bf x}_{2})d{\bf x}_{2},
\label{e6}
\end{eqnarray}
\begin{eqnarray}
\label{e7} {\partial g\over\partial t}+{\bf v}_{1}{\partial g\over\partial {\bf r}_{1}}+\langle {\bf F}\rangle_{1} {\partial g\over\partial {\bf v}_{1}}+\frac{1}{m^{2}}
 {\bf {\cal F}}(2\rightarrow 1)f_{2}  {\partial f_{1}\over\partial {\bf v}_{1}}\nonumber\\
+\frac{\partial}{\partial {\bf v}_{1}}\int {\bf F}(3\rightarrow 1)g({\bf x}_{2},{\bf x}_{3},t) \frac{f_{1}}{m}d{\bf x}_{3}+(1\leftrightarrow 2)=0,
\end{eqnarray}
where we have introduced the abbreviations $f_{1}=f({\bf r}_{1},{\bf
v}_{1},t)$ and $f_{2}=f({\bf r}_{2},{\bf v}_{2},t)$. We have also
introduced the  mean force (by unit of mass) created in ${\bf r}_1$ by all the
particles
\begin{eqnarray}
\label{e8}
\langle {\bf F}\rangle_{1} =\int {\bf F}(2\rightarrow 1)\frac{f_2}{m}d{\bf r}_{2}d{\bf v}_{2}=-\nabla\Phi_{1},
\end{eqnarray}
and  the fluctuating force (by unit of mass) created by particle $2$
on particle $1$:
\begin{eqnarray}
\label{e9}
{\bf {\cal F}}(2\rightarrow 1)={\bf F}(2\rightarrow
1)-\frac{1}{N}\langle {\bf F}\rangle_{1}.
\end{eqnarray}
These equations are exact at the order $O(1/N)$ where the three-body
correlation function can be neglected. They form therefore the right
basis to develop a kinetic theory for Hamiltonian systems with weak
long-range interactions. We note that these equations are similar to
the BBGKY hierarchy of plasma physics but not identical. One
difference is the $(N-1)/N$ term in Eq. (\ref{e6}). The other
difference is the presence of the fluctuating force ${\cal
F}(2\rightarrow 1)$ instead of $F(2\rightarrow 1)$ due to the spatial
inhomogeneity of the system. In plasma physics, the system is
homogeneous over distances of the order of the Debye length so the
mean force $\langle {\bf F}\rangle$ vanishes.

\subsection{3.2. The limit $N\rightarrow +\infty$: the Vlasov equation}

Recalling that $P_{2}'\sim 1/N$, we note that
\begin{eqnarray}
P_{2}({\bf x}_{1},{\bf x}_{2},t)=P_{1}({\bf
x}_{1},t) P_{1}({\bf x}_{2},t)+O(1/N).
\label{v1}
\end{eqnarray}
If we consider the limit $N\rightarrow +\infty$ (for a fixed time
$t$), we see that the correlations between particles can be neglected
so the two-body distribution function factorizes in two one-body
distribution functions i.e. $P_{2}({\bf x}_{1},{\bf
x}_{2},t)=P_{1}({\bf x}_{1},t) P_{1}({\bf x}_{2},t)$. Therefore the
{\it mean field approximation} is exact in the limit $N\rightarrow
+\infty$. Substituting this result in Eq. (\ref{h1}), we obtain the
Vlasov equation
\begin{eqnarray}
{\partial f_{1}\over\partial t}+{\bf v}_{1}{\partial f_{1}\over\partial
{\bf r}_{1}}+\langle {\bf F}\rangle_{1} {\partial f_{1}\over \partial
{\bf v}_{1}}=0. \label{v2}
\end{eqnarray}
This equation also results from Eq. (\ref{e6}) if we neglect the
correlation function $g=N^{2}P_{2}'$ in the r.h.s. The Vlasov equation
describes the {\it collisionless evolution} of the system up to a time
at least of order $N t_D$ (where $t_D$ is the dynamical time). In
practice, $N\gg 1$ so that the domain of validity of the Vlasov
equation is huge (for example, in stellar systems $N\sim 10^6-10^{12}$
stars). When the Vlasov equation is coupled to a long-range potential
of interaction it can develop a process of phase mixing and violent
relaxation leading to a quasi-stationary state (QSS) on a very short
timescale, of the order of the dynamical time $t_{D}$ \cite{lb}. This
process will be discussed in Sec. 4.

\subsection{3.3. The order $O(1/N)$: a general kinetic equation}

If we want to
describe the collisional evolution of the system, we need to consider
finite $N$ effects.  Equations (\ref{e6}) and (\ref{e7}) describe the
evolution of the system on a timescale of order $N t_D$. The equation
for the evolution of the smooth distribution function is of the form
\begin{eqnarray}
{\partial f_{1}\over\partial t}+{\bf v}_{1}{\partial f_{1}\over\partial
{\bf r}_{1}}+\frac{N-1}{N}\langle {\bf F}\rangle_{1} {\partial
f_{1}\over \partial {\bf v}_{1}}=C_{N}[f_{1}], \label{e11}
\end{eqnarray}
where $C_{N}$ is a ``collision'' term analogous to the one arising in
the Boltzmann equation. In the present context, there are not real
collisions between particles. The term on the right hand side of Eq.
(\ref{e11}) is due to the development of correlations between
particles as time goes on. It is related to the two-body correlation
function $g({\bf x}_{1},{\bf x}_{2},t)$ which is itself related to the
distribution function $f({\bf x}_{1},t)$ by Eq. (\ref{e7}). Our aim is
to obtain an expression for the collision term $C_{N}[f]$ at the order
$1/N$. The difficulty with Eq. (\ref{e7}) for the two-body correlation
function is that it is an integrodifferential equation. The second and
third terms are advective terms, the fourth term is the {\it source}
of the correlation and the fifth term takes into account {\it
collective effects}. If we neglect the contribution of the integral in
Eq. (\ref{e7}), we get the simpler coupled system
\begin{eqnarray}
{\partial f_{1}\over\partial t}+{\bf v}_{1}{\partial f_{1}\over\partial {\bf r}_{1}}+\frac{N-1}{N}\langle {\bf F}\rangle_{1} {\partial f_{1}\over \partial {\bf v}_{1}}=-m {\partial \over\partial {\bf v}_{1}}\int {\bf F}(2\rightarrow 1)g({\bf x}_{1},{\bf x}_{2})d{\bf x}_{2},
\label{e12}
\end{eqnarray}
\begin{eqnarray}
\label{e13} {\partial g\over\partial t}+\left \lbrack {\bf v}_{1}{\partial\over\partial {\bf r}_{1}}+{\bf v}_{2}{\partial\over\partial {\bf r}_{2}} +\langle {\bf F}\rangle_{1} {\partial\over\partial {\bf v}_{1}}+\langle {\bf F}\rangle_{2} {\partial\over\partial {\bf v}_{2}}\right  \rbrack g\nonumber\\
+\left \lbrack {\bf {\cal F}}(2\rightarrow 1) {\partial\over\partial {\bf v}_{1}}+{\bf {\cal F}}(1\rightarrow 2) {\partial\over\partial {\bf v}_{2}}\right \rbrack \frac{f_{1}}{m}\frac{f_{2}}{m}=0.
\end{eqnarray}
The integral that we have neglected contains ``collective effects''
that describe the polarization of the medium. In plasma physics, they
are responsible for the Debye shielding. These collective effects are
taken into account in the Lenard-Balescu equation through the
dielectric function \cite{balescubook}. However, this equation is
restricted to spatially homogeneous systems and based on a Markovian
approximation (see, e.g.,
\cite{paper2}). These assumptions are necessary to use Laplace-Fourier
transforms in order to solve the integro-differential equation
(\ref{e7}). Here, we want to describe more general situations where
the interaction is not shielded so that the system can be spatially
{\it inhomogeneous}. If we neglect collective effects, we can obtain a
general kinetic equation in a closed form (\ref{general}) that is
valid for systems that are not necessarily homogeneous and that can
take into account memory effects. This equation has interest in its
own right (despite its limitations) because its structure bears a lot
of physical significance.

The equation (\ref{e13}) for the correlation function can be
written
\begin{eqnarray}
\label{g1a} {\partial g\over\partial t}+{\cal L} g=-\left \lbrack {\bf {\cal F}}(2\rightarrow 1) {\partial\over\partial {\bf v}_{1}}+{\bf {\cal F}}(1\rightarrow 2) {\partial\over\partial {\bf v}_{2}}\right \rbrack \frac{f}{m}({\bf x}_{1},t)\frac{f}{m}({\bf x}_{2},t),
\end{eqnarray}
where we have denoted the advective term by ${\cal L}$ (Liouvillian
operator). Solving formally this equation with the Green function
\begin{eqnarray}
G(t,t')={\rm exp}\left\lbrace -\int_{t'}^t {\cal
L}(\tau)d\tau\right\rbrace,\label{g2yrt}
\end{eqnarray}
we obtain
\begin{eqnarray}
g({\bf x}_1,{\bf x}_2,t)=-\int_0^t d\tau G(t,t-\tau) \left \lbrack {\bf {\cal F}}(2\rightarrow 1) {\partial\over\partial {\bf v}_{1}}+{\bf {\cal F}}(1\rightarrow 2) {\partial\over\partial {\bf v}_{2}}\right \rbrack\frac{f}{m}({\bf
x}_1,t-\tau)\frac{f}{m}({\bf x}_2,t-\tau).\nonumber\\
\label{gen3}
\end{eqnarray}
The Green function constructed with the smooth field $\langle {\bf
F}\rangle$ means that, in order to evaluate the time integral in Eq.
(\ref{gen3}), we must move the coordinates ${\bf r}_i(t-\tau)$ and ${\bf
v}_i(t-\tau)$ of the particles with the mean field flow in phase
space, adopting a Lagrangian point of view. Thus, in evaluating the
time integral, the coordinates ${\bf r}_{i}$ and ${\bf v}_{i}$ placed
after the Greenian must be viewed as ${\bf r}_{i}(t-\tau)$ and ${\bf
v}_{i}(t-\tau)$ where
\begin{eqnarray}
{\bf r}_{i}(t-\tau)={\bf r}_{i}(t)-\int_{0}^{\tau}{\bf v}_{i}(t-s)ds, \quad {\bf v}_{i}(t-\tau)={\bf v}_{i}(t)-\int_{0}^{\tau}\langle {\bf F}\rangle ({\bf r}_{i}(t-s),t-s)ds.
\label{g3c}
\end{eqnarray}
Substituting Eq. (\ref{gen3}) in Eq. (\ref{e12}), we get
\begin{eqnarray}
\frac{\partial f_1}{\partial t}+{\bf v}_{1}{\partial f_{1}\over\partial {\bf r}_{1}}+\frac{N-1}{N}\langle {\bf F}\rangle_{1} {\partial f_{1}\over \partial {\bf v}_{1}}=\frac{\partial}{\partial
{v}_1^{\mu}}\int_0^t d\tau \int d{\bf r}_{2}d{\bf v}_2
{F}^{\mu}(2\rightarrow
1,t)G(t,t-\tau)\nonumber\\
\times  \left \lbrack {{\cal F}}^{\nu}(2\rightarrow 1) {\partial\over\partial { v}_{1}^{\nu}}+{{\cal F}}^{\nu}(1\rightarrow 2) {\partial\over\partial {v}_{2}^{\nu}}\right \rbrack {f}({\bf r}_1,{\bf v}_1,t-\tau)\frac{f}{m}({\bf r}_2,{\bf
v}_2,t-\tau). \label{general}
\end{eqnarray}
This kinetic equation can also be obtained from a more abstract
projection operator formalism \cite{kandrup1} or from a quasilinear
theory based on the Klimontovich equation \cite{paper4} (note that we
can replace ${F}^{\mu}(2\rightarrow 1,t)$ by ${\cal
F}^{\mu}(2\rightarrow 1,t)$ in the first term of the r.h.s. of the
equation since the fluctuations vanish in average).  This kinetic
equation (\ref{general}) is valid at order $1/N$ so that it describes
the ``collisional'' evolution of the system on a timescale of order
$Nt_D$ (ignoring collective effects). Equation (\ref{general}) is a
non-Markovian integrodifferential equation. If we implement a
Markovian approximation ${f}({\bf r}_1,{\bf v}_1,t-\tau)\simeq
{f}({\bf r}_1,{\bf v}_1,t)$, ${f}({\bf r}_2,{\bf v}_2,t-\tau)\simeq
{f}({\bf r}_2,{\bf v}_2,t)$ and extend the time integral to $+\infty$,
we obtain
\begin{eqnarray}
\frac{\partial f_1}{\partial t}+{\bf v}_{1}{\partial f_{1}\over\partial {\bf r}_{1}}+\frac{N-1}{N}\langle {\bf F}\rangle_{1} {\partial f_{1}\over \partial {\bf v}_{1}}=\frac{\partial}{\partial
{v}_1^{\mu}}\int_0^{+\infty} d\tau \int d{\bf r}_{2}d{\bf v}_2
{F}^{\mu}(2\rightarrow
1,t)G(t,t-\tau)\nonumber\\
\times  \left \lbrack {{\cal F}}^{\nu}(2\rightarrow 1) {\partial\over\partial { v}_{1}^{\nu}}+{{\cal F}}^{\nu}(1\rightarrow 2) {\partial\over\partial {v}_{2}^{\nu}}\right \rbrack {f}({\bf r}_1,{\bf v}_1,t)\frac{f}{m}({\bf r}_2,{\bf
v}_2,t). \label{gm}
\end{eqnarray}
The Markov approximation is justified for $N\rightarrow +\infty$
because the timescale $Nt_{D}$ on which $f$ changes is long compared
with the timescale $\tau_{corr}$ for which the integrand in
Eq. (\ref{gm}) has significant support (except for self-gravitating
systems for which the force auto-correlation function decreases like
$1/t$ \cite{ct}). We do not assume, however, that the
decorrelation time is ``extremely'' short (i.e,
$\tau_{corr}\rightarrow 0$). Therefore, in the time integral, the
distribution functions must be evaluated at $({\bf r}(t-\tau),{\bf
v}(t-\tau))$ and $({\bf r}_{1}(t-\tau),{\bf v}_{1}(t-\tau))$ where
\begin{eqnarray}
{\bf r}_{i}(t-\tau)={\bf r}_{i}(t)-\int_{0}^{\tau}{\bf v}_{i}(t-s)ds,\quad 
{\bf v}_{i}(t-\tau)={\bf v}_{i}(t)-\int_{0}^{\tau}\langle {\bf F}\rangle ({\bf r}_{i}(t-s),t)ds.
\label{g3cq}
\end{eqnarray}
Comparing Eq. (\ref{g3cq}) with Eq. (\ref{g3c}), we have assumed that
the mean force $\langle{\bf F}\rangle ({\bf r},t)$ does not change
substantially on the timescale $\tau_{corr}$ on which the time
integral has essential contributions.

\subsection{3.4. Properties of the general kinetic equation}

When the system is spatially inhomogeneous, its collisional evolution
can be very complicated and very little is known concerning kinetic
equations of the form (\ref{general}). For example, it is not
straightforward to prove by a direct calculation that
Eq. (\ref{general}) conserves the energy. However, since
Eq. (\ref{general}) is exact at order $O(1/N)$, the energy must be
conserved. Indeed, the integral constraints of the Hamiltonian system
must be conserved at any order of the $1/N$ expansion \footnote{As we
have already indicated, Eq. (\ref{general}) ignores collective
effects. However, we do not think that this approximation alters the
conservation of the energy. When we make a Markovian approximation
(valid for $N\rightarrow +\infty$) and extend the time integral to
infinity, leading to Eq. (\ref{gm}), the conservation of the energy is
shown in Sec. 3.5 for homogeneous systems and in \cite{angle} for
inhomogeneous systems (using angle-action variables). }.  On the other
hand, we cannot establish the $H$-theorem for an equation of the form
(\ref{general}). It is only when additional approximations are
implemented (markovian approximation) that the $H$-theorem is
obtained. To be more precise, let us compute the rate of change of the
Boltzmann entropy $S_B=-\int
\frac{f_1}{m}\ln \frac{f_1}{m} d{\bf r}_1 d{\bf v}_1$ with respect to
the general kinetic equation (\ref{general}). After straightforward
manipulations obtained by interchanging the indices $1$ and $2$, it
can be put in the form
\begin{eqnarray}
\dot S_{B}=\frac{1}{2m^{2}}\int d{\bf x}_{1}d{\bf x}_{2}\frac{1}{f_{1}f_{2}}\int_{0}^{t}d\tau \left\lbrack {\cal F}^{\mu}(2\rightarrow 1)f_2\frac{\partial f_{1}}{\partial v_{1}^{\mu}}+{\cal F}^{\mu}(1\rightarrow 2)f_1\frac{\partial f_{2}}{\partial v_{2}^{\mu}}\right\rbrack_{t} \nonumber\\
\times G(t,t-\tau)\left\lbrack {\cal F}^{\nu}(2\rightarrow 1)f_2\frac{\partial f_{1}}{\partial v_{1}^{\nu}}+{\cal F}^{\nu}(1\rightarrow 2)f_1\frac{\partial f_{2}}{\partial v_{2}^{\nu}}\right\rbrack_{t-\tau}.\label{g6fsef}
\end{eqnarray}
We note that its sign is not necessarily positive. This depends on the
importance of memory effects. In general, the Markovian approximation
is justified for $N\rightarrow +\infty$ because the correlations decay
on a timescale $\tau_{corr}$ that is much smaller than the time $\sim
Nt_{D}$ on which the distribution changes (with, again, the exception
of gravity). In that case, the entropy increases monotonically as
shown in Sec. 3.5 for homogeneous systems and in \cite{angle} for
inhomogeneous systems (using angle-action variables). Now, even if the
energy is conserved and the entropy increases monotonically, it is not
completely clear whether the general kinetic equation (\ref{general})
will relax towards the mean field Maxwell-Boltzmann distribution
(\ref{bol9}) of statistical equilibrium. It could be trapped in a
steady state that is not the state of maximal entropy because there is
no resonance anymore to drive the relaxation (see the discussion in
Secs. 3.5 and 3.7 and in \cite{angle}). Indeed, the kinetic equation
(\ref{general}) can admit several stationary solutions, not only the
Boltzmann distribution. It could also undergo everlasting oscillations
without reaching a steady state.  The kinetic equation (\ref{general})
may have a rich variety of behaviors and its complete study is of
great complexity.

\subsection{3.5. Spatially homogeneous systems: the Landau and Lenard-Balescu equations}

If we consider spatially homogeneous systems, make a Markovian
approximation and extend the time integration to infinity, the kinetic
equation (\ref{general}) becomes
\begin{eqnarray}
\frac{\partial f_1}{\partial t}=\frac{\partial}{\partial
{v}_1^{\mu}}\int_0^{+\infty} d\tau \int d{\bf r}_{2}d{\bf v}_2
{F}^{\mu}(2\rightarrow
1,t) {F}^{\nu}(2\rightarrow 1,t-\tau) \nonumber\\
\times \left ( {\partial\over\partial { v}_{1}^{\nu}}-{\partial\over\partial {v}_{2}^{\nu}}\right ) {f}({\bf v}_1,t)\frac{f}{m}({\bf
v}_2,t), \label{hom1}
\end{eqnarray}
where we have used ${\bf F}(2\rightarrow 1)=-{\bf F}(1\rightarrow 2)$.
Since $\langle {\bf F}\rangle={\bf 0}$, the particles follow linear
trajectories with constant velocity to leading order in
$N$. Therefore, ${\bf v}_{i}(t-\tau)={\bf v}_{i}$ and ${\bf
r}_{i}(t-\tau)={\bf r}_{i}-{\bf v}_{i}\tau$ where ${\bf r}_{i}$ and
${\bf v}_{i}$ denote their position and velocity at time $t$.  In that
case, the integrals on time and position in Eq. (\ref{hom1}) can be easily
performed in Fourier space \cite{paper2} and we obtain the Landau equation
\begin{eqnarray}
\frac{\partial f_1}{\partial t}=\pi (2\pi)^d m
\frac{\partial}{\partial {v}_1^{\mu}} \int d{\bf v}_2 d{\bf k}
k^{\mu}k^{\nu} \hat{u}(k)^2 \delta({\bf k}\cdot {\bf w}) \left
(f_2\frac{\partial f_1}{\partial v_1^{\nu}}-f_1\frac{\partial
f_2}{\partial v_2^{\nu}}\right ), \label{hom2}
\end{eqnarray}
where ${\bf w}={\bf v}_{1}-{\bf v}_{2}$. Other equivalent expressions
of the Landau equation are given in \cite{paper2}. The Landau equation
ignores collective effects. Collective effects can be taken into
account by keeping the contribution of the last integral in
Eq. (\ref{e7}). For spatially homogeneous systems, the calculations
can be carried out explicitly in the complex plane
\cite{ichimaru} and lead to the Lenard-Balescu equation 
\begin{eqnarray}
\frac{\partial f_1}{\partial t}=\pi (2\pi)^d m
\frac{\partial}{\partial {v}_1^{\mu}} \int d{\bf v}_2 d{\bf k}
k^{\mu}k^{\nu} \frac{\hat{u}(k)^2}{|\epsilon({\bf k},{\bf k}\cdot {\bf v}_{2})|^{2}} \delta({\bf k}\cdot {\bf w}) \left
(f_2\frac{\partial f_1}{\partial v_1^{\nu}}-f_1\frac{\partial
f_2}{\partial v_2^{\nu}}\right ), \label{hom3}
\end{eqnarray}
where $\epsilon({\bf k},\omega)$ is the dielectric function
\begin{eqnarray}
\epsilon({\bf k},\omega)=1-(2\pi)^{d}\hat{u}({\bf k})\int \frac{{\bf k}\cdot \partial f/\partial {\bf v}}{{\bf k}\cdot {\bf v}-\omega}d{\bf v}. \label{hom4}
\end{eqnarray}
We note that the Lenard-Balescu equation is obtained from the
Landau equation by replacing the potential $\hat{u}(k)$ by the
``screened'' potential $\hat{u}(k)/|\epsilon({\bf k},{\bf k}\cdot
{\bf v}_{2})|$. The Landau and Lenard-Balescu equations conserve mass
and energy (reducing to the kinetic energy for a spatially homogemous
system) and monotonically increase the Boltzmann entropy
\cite{balescubook}. Thus, $\dot M=\dot E=0$ and $\dot S_{B}\ge 0$ 
($H$-theorem). The collisional evolution is due to a condition of
resonance between the particles' orbits. For homogeneous systems, the
condition of resonance encapsulated in the $\delta$-function appearing
in the Landau and Lenard-Balescu equations corresponds to ${\bf
k}\cdot {\bf v}_{1}={\bf k}\cdot {\bf v}_{2}$ with ${\bf v}_{1}\neq
{\bf v}_{2}$. For $d>1$ the only stationary solution is the Maxwell
distribution. Therefore, the Landau and Lenard-Balescu equations relax
towards the Maxwell distribution.  Since the collision term in
Eq. (\ref{general}) is valid at order $O(1/N)$, the relaxation time scales
like
\begin{eqnarray}
t_{R}\sim Nt_D, \qquad (d>1)
 \label{l8yhy}
\end{eqnarray}
as can be seen directly from Eq. (\ref{hom2}) by dimensional analysis
(comparing the l.h.s. and the r.h.s., we have $1/t_{R}\sim
u_{*}^{2}N\sim 1/N$ while $t_{D}\sim R/v_{typ}\sim 1$ with the
scalings introduced in Sec. 2.2). This scaling, predicted in
\cite{paper2}, has been observed for 2D Coulombian plasmas
\cite{benedetti,landaud}.

For one-dimensional systems, like the HMF model, the situation is
different. For $d=1$, the kinetic equation (\ref{hom2}) reduces to
\begin{eqnarray}
\frac{\partial f_1}{\partial t}=2\pi^2 m
\frac{\partial}{\partial {v}_1} \int d{v}_2 d{k}
{k^{2}\over |k|} \frac{\hat{u}(k)^2}{|\epsilon(k,k v_{2})|^{2}} \delta(v_{1}-v_{2}) \left
(f_2\frac{\partial f_1}{\partial v_1}-f_1\frac{\partial
f_2}{\partial v_2}\right )=0. \label{hom5}
\end{eqnarray}
Therefore, the collision term $C_{N}[f]$ vanishes at the order $1/N$
because there is no resonance. The kinetic equation reduces to
$\partial f/\partial t=0$ so that the distribution function does not
evolve at all on a timescale $\sim Nt_{D}$. This implies that, for
one-dimensional homogeneous systems, the relaxation time towards statistical
equilibrium is larger than $Nt_D$. Thus, we expect that
\begin{eqnarray}
t_{R}> N t_D, \qquad (d=1).
 \label{l8yh}
\end{eqnarray}
For the HMF model, when the system remains spatially homogeneous, it
is found that the relaxation time scales like $t_{R}\sim e^{N}$
\cite{campa}. The fact that the Lenard-Balescu collision term vanishes
in 1D is known for a long time in plasma physics (see, e.g., the last
paragraph in \cite{kp}) and has been rediscovered recently in the
context of the HMF model \cite{bd,cvb}.

\subsection{3.6. The Vlasov-Landau equation for stellar systems}

Self-gravitating systems are spatially inhomogeneous, but the
collisional current can be calculated by making a {\it local
approximation} \cite{bt}. Therefore, the evolution of the distribution
function is governed by the Vlasov-Landau equation 
\begin{equation}
\label{bol13} {\partial f_{1}\over\partial t}+{\bf v}_{1}\cdot
{\partial f_{1}\over\partial {\bf r}_{1}}+\langle {\bf F}\rangle_{1}\cdot {\partial
f_{1}\over\partial {\bf v}_{1}}={\partial\over\partial
v_{1}^{\mu}}\int  K^{\mu\nu}\biggl (f_{2}{\partial
f_{1}\over\partial v_{1}^{\nu}}-f_{1}{\partial
f_{2}\over\partial v_{2}^{\nu}}\biggr )d{\bf v}_{2},
\end{equation}
\begin{equation}
\label{bol14}
K^{\mu\nu}=2\pi m G^{2}{1\over w}\ln\Lambda \biggl (\delta^{\mu\nu}-{w^{\mu}w^{\nu}\over w^{2}}\biggr ),
\end{equation}
where $\ln
\Lambda=\int_{k_{min}}^{k_{max}}dk/k$ is the Coulomb factor (regularized
with appropriate cut-offs) and we have set $f_1=f({\bf r}_{1},{\bf
v}_{1},t)$ and $f_2=f({\bf r}_{1},{\bf v}_{2},t)$ assuming that the
encounters can be treated as local (see, e.g.,
\cite{kandrup1,saslaw,spitzer,paper3} for a more complete discussion). The Vlasov-Landau-Poisson system
conserves the total mass and the total energy (kinetic $+$ potential)
of the system. It also increases the Boltzmann entropy (\ref{bol4})
monotonically: $\dot S_{B}\ge 0$ (H-theorem). The mean field
Maxwell-Boltzmann distribution (\ref{bol9}) is the only stationary
solution of this equation (cancelling both the advective term and the
collision term individually). Therefore, the system {\it tends} to
reach the Boltzmann distribution.  However, there are two reasons why
it cannot attain it: (i) {\it Evaporation:} when coupled to the
gravitational Poisson equation, the mean field Maxwell-Boltzmann
distribution (\ref{bol9}) yields a density profile $\rho\sim r^{-2}$
(for $r\rightarrow +\infty$) with infinite mass so there is no
physical distribution of the form (\ref{bol9}) in an infinite domain
\cite{paddy,ijmpb}.  The system can increase the Boltzmann entropy
indefinitely by evaporating. Therefore, the Vlasov-Landau-Poisson
system has no steady state with finite mass and the density profile
tends to spread indefinitely. (ii) {\it Gravothermal catastrophe:} if
the energy of the system is lower than the Antonov threshold
$E_{c}=-0.335GM^{2}/R$ (where $R$ is the system size), it will undergo
core collapse. This is called gravothermal catastrophe \cite{lbw}
because the system can increase the Boltzmann entropy indefinitely by
contracting and overheating. This process usually dominates over
evaporation and leads to the formation of binary stars
\cite{bt,heggie,ijmpb}.  If the system is confined within a box so as
to prevent evaporation, the Vlasov-Landau-Poisson system admits
stationary solutions for sufficiently large energies (above the
Antonov threshold). They correspond to the mean field Boltzmann
distributions (\ref{bol9}). There can exist several Boltzmann
distributions with the same values of mass and energy having different
density constrasts ${\cal R}=\rho(0)/\rho(R)$. Since the Boltzmann
entropy is the Lyapunov functional of the Vlasov-Landau equation, a
Boltzmann distribution (steady state) is linearly dynamically stable
iff it is a (local) maximum of entropy at fixed mass and energy
(thermodynamical stability).  Therefore, only the local entropy maxima
at fixed mass and energy can be reached in practice. This corresponds
to configurations with density contrast ${\cal R}<709$. Minima and
saddle points of entropy (${\cal R}>709$) are unstable. The relaxation
time to the Boltzmann distribution is
\begin{eqnarray}
t_{R}\sim  \frac{N}{\ln N} t_D, 
 \label{l8yhb}
\end{eqnarray}
where the logarithmic correction comes from the divergence of the
Coulombian factor $\ln\Lambda\sim \ln N$. Since there is no global
entropy maximum at fixed mass and energy for a self-gravitating system
confined within a box \cite{paddy,ijmpb}, the equilibrium
configurations are only {\it metastable} (local entropy
maxima). However, the lifetime of the metastable states is
considerable, scaling like $e^{N}$, so that these states are stable in
practice \cite{lifetime}. Depending on the initial condition and on
the structure of the basin of attraction, the system can either relax
towards a long-lived metastable mean field Boltzmann distribution or
undergo gravitational collapse (gravothermal catastrophe) and form a
binary star surrounded by a hot halo (leading to a structure with
unbounded entropy).

\subsection{3.7. Orbit-averaged kinetic equation}

We have seen in
Sec. 3.5 that one dimensional systems that are spatially
homogeneous do not evolve at all on a timescale $\sim Nt_{D}$ or
larger because of the absence of resonances. However, if the system is
spatially inhomogeneous, new resonances can appear as described in
\cite{angle} so that an evolution is possible on a timescale
$Nt_{D}$. Then, we can expect that one dimensional inhomogeneous
systems will {\it tend} to approach the Boltzmann distribution on the
timescale $Nt_{D}$.  To be
more precise, let us consider the orbit-averaged-Fokker-Planck
equation derived in
\cite{angle}. Exploiting the timescale separation between the
dynamical time and the relaxation time, we can average Eq. (\ref{gm})
over the orbits, assuming that at any stage of its evolution the
system reaches a mechanical equilibrium on a short dynamical
time. Therefore, the distribution function is a quasi-stationary solution of
the Vlasov equation $f\simeq f(\epsilon,t)$ [where
$\epsilon=v^{2}/2+\Phi$ is the individual energy] slowly evolving in
time under the effect of ``collisions'' ($=$ correlations due to
finite $N$ effects). Introducing angle-action variables, we get an
equation of the form \cite{angle}:
\begin{equation}
\label{q16}
\frac{\partial f}{\partial t}=\frac{1}{2}\frac{\partial}{\partial J}\sum_{m,m'}\int {m |A_{mm'}(J,J')|^{2}}\delta(m\Omega(J)-m'\Omega(J'))\left\lbrace f(J')m\frac{\partial f}{\partial J}-f(J)m'\frac{\partial f}{\partial J'}\right\rbrace dJ'.
\end{equation} 
The important point to notice is that the evolution of the system is
due to a condition of resonance between the pulsations $\Omega(J)$ of
the particles' orbits (this property extends to $d$ dimensions). Only
particles whose pulsations satisfy $m\Omega(J)=m'\Omega(J')$ with
$(m,J)\neq (m',J')$ participate to the diffusion current. This is
similar to the collisional relaxation of two dimensional point
vortices
\cite{kinvortex,bbgky}. It can be shown that Eq. (\ref{q16})
conserves mass and energy and monotonically increases the Boltzmann
entropy so that the system {\it tends} to approach the Boltzmann
distribution of statistical equilibrium on a timescale $\sim Nt_{D}$
(the Boltzmann distribution is always a stationary solution of
Eq. (\ref{q16})) \cite{angle}. However, it may happen that there is
not enough resonances so that the system can be trapped in a quasi
stationary state {\it different} from the Boltzmann distribution. This
happens when the condition of resonance cannot be satisfied so that
$m\Omega(J)\neq m'\Omega(J')$ for all $(m,J)\neq (m',J')$ for which
$|A_{mm'}(J,J')|^{2}\neq 0$. In that case, the system is in a steady
state of Eq. (\ref{q16}) that is not the Boltzmann
distribution. Indeed, Eq. (\ref{q16}) may admit other stationary
solutions than the Boltzmann distribution \cite{angle}. This is what
happens to point vortices in 2D hydrodynamics: the collisional
relaxation stops when the profile of angular velocity becomes
monotonic even if the system has not reached the Boltzmann
distribution
\cite{kinvortex}.  In that case, the system will relax towards the Boltzmann distribution (if it truly does!)  on
a timescale larger than $Nt_{D}$ (this requires to develop the kinetic
theory at higher order in $1/N$, taking into account three body, four
body,... correlations). We may wonder whether the same situation can
happen to systems described by a kinetic equation of the form
(\ref{general}).

\section{4. Statistical theory of violent relaxation: the Lynden-Bell entropy}

For $t\ll Nt_{D}$, the evolution of the system  is
governed by the Vlasov equation
\begin{equation}
\label{vp1}
{\partial f\over\partial t}+{\bf v}\cdot {\partial f\over\partial {\bf r}}+{\bf F}\cdot {\partial f\over\partial {\bf v}}=0,
\end{equation}
where ${\bf F}=-\nabla\Phi$ is the force by unit of mass
experienced by a particle. It is produced self-consistently by the other particles through the mean field
\begin{equation}
\label{vp2}
\Phi({\bf r},t)=\int u({\bf r}-{\bf r}')\rho({\bf r}',t)d{\bf r}'.
\end{equation}
Mathematically, the Vlasov equation is obtained when the $N\rightarrow
+\infty$ limit is taken before the $t\rightarrow +\infty$
limit. Indeed, the collision term in Eq.  (\ref{general}) scales as
$1/N$ so that it vanishes for $N\rightarrow +\infty$. The Vlasov
equation, or collisionless Boltzmann equation, simply states that, in
the absence of ``collisions'' (more properly correlations), the
distribution function $f$ is conserved by the flow in phase
space. This can be written $df/dt=0$ where $d/dt=\partial/\partial
t+{\bf v}\cdot \partial/\partial {\bf
r}-\nabla\Phi\cdot\partial/\partial {\bf v}$ is the material
derivative.  The Vlasov equation conserves the usual constraints: the
mass $M$ and the energy $E$, but also an infinite number of invariants
called the Casimirs. They are defined by $ I_{h}=\int h(f) d{\bf r}
d{\bf v}$ for any continuous function $h(f)$. The conservation of the
Casimirs is equivalent to the conservation of the moments of the
distribution function denoted
\begin{equation}
\label{vp3} M_n=\int f^{n} d{\bf r}d{\bf v}.
\end{equation}

The Vlasov equation admits an infinite number of stationary solutions
whose general form is given by the Jeans theorem \cite{bt}.
Starting from an unstable initial condition $f_{0}({\bf r},{\bf v})$,
the Vlasov equation develops very complex filaments as a result
of a mixing process in phase space (collisionless mixing).  In this
sense, the fine-grained distribution function ${f}({\bf r},{\bf v},t)$
will never reach a stationary state but will rather produce
intermingled filaments at smaller and smaller scales.  However, if we
introduce a coarse-graining procedure, the coarse-grained distribution
function $\overline{f}({\bf r},{\bf v},t)$ will reach a
metaequilibrium state $\overline{f}({\bf r},{\bf v})$ on a very short
timescale, of the order of the dynamical time $t_D$. This is because
the evolution continues at scales smaller than the scale of
observation (coarse-grained).  This process is known as ``phase
mixing'' and ``violent relaxation'' (or collisionless relaxation)
\cite{bt}. Then, on a longer timescale, collisions (correlations) will  come into play and drive the system towards the Boltzmann 
statistical equilibrium state.  Lynden-Bell
\cite{lb} has tried to predict the metaequilibrium state, or quasi stationary state (QSS), achieved
by the system through violent relaxation in terms of statistical
mechanics. This approach is of course quite distinct from the
statistical mechanics of the $N$-body system (exposed in Sec. 2.4)
which describes the statistical equilibrium state reached by a
discrete $N$-body Hamiltonian system for $t\rightarrow +\infty$.  In
Lynden-Bell's approach, we make the statistical mechanics of a {\it
field}, the distribution function $f({\bf r},{\bf v},t)$ whose
evolution is governed by the Vlasov equation (\ref{vp1}) while in Sec.
2.4 we made the statistical mechanics of a system of {\it
point} particles described by the Hamilton equations (\ref{eqb1}).

Let $f_{0}({\bf r},{\bf v})$ denote the initial (fine-grained)
distribution function (DF). We discretize $f_{0}({\bf r},{\bf v})$ in
a series of levels $\eta$ on which $f_{0}({\bf r},{\bf v})\simeq \eta$
is approximately constant. Thus, the levels $\lbrace \eta\rbrace$
represent all the values taken by the fine-grained DF. If the initial
condition is unstable, the distribution function $f({\bf r},{\bf
v},t)$ will be stirred in phase space (phase mixing) but will conserve
its values $\eta$ and the corresponding hypervolumes
$\gamma(\eta)=\int \delta(f({\bf r},{\bf v},t)-\eta)d{\bf r}d{\bf v}$
as a property of the Vlasov equation (this is equivalent to the
conservation of the Casimirs). Let us introduce the probability
density $\rho({\bf r},{\bf v},\eta)$ of finding the level of phase
density $\eta$ in a small neighborhood of the position ${\bf r},{\bf
v}$ in phase space. This probability density can be viewed as the
local area proportion occupied by the phase level $\eta$ and it must
satisfy at each point the normalization condition
\begin{equation}
\label{E1}
\int\rho({\bf  r,v},\eta)d\eta=1.
\end{equation}
The locally averaged (coarse-grained) DF is
then expressed in terms of the probability density as
\begin{equation}
\label{E2}
\overline{f}({\bf  r,v})=\int\rho({\bf  r,v}, \eta)\eta d \eta,
\end{equation}
and  the associated macroscopic potential satisfies
\begin{equation}
\label{vp2b}
\overline{\Phi}({\bf r},t)=\int u({\bf r}-{\bf r}')\overline{f}({\bf r}',{\bf v}',t)d{\bf r}'d{\bf v}'.
\end{equation}
Since the potential is expressed by space integrals of
the density, it smoothes out the fluctuations of the distribution
function, supposed at very fine scale, so $\Phi$ has negligible
fluctuations (we thus drop the bar on $\Phi$). The conserved
quantities of the Vlasov equation can be decomposed in two groups. The
mass and energy will be called {\it robust integrals} because they are
conserved by the  coarse-grained distribution function:
$\overline{M[f]}=M[\overline{f}]$ and $\overline{E[f]}\simeq
E[\overline{f}]$. Hence
\begin{equation}
\label{E4}
M=\int \overline{f}d{\bf  r} d{\bf  v},
\end{equation}
\begin{equation}
\label{E5}
E=\int{1\over 2}\overline{f} v^{2} d{\bf  r}
 d{\bf  v}+{1\over 2}\int \overline{f} \ {\Phi} d{\bf  r} d{\bf  v}.
\end{equation}
As discussed above, the potential can be considered
as smooth, so we have expressed the energy in terms of the
coarse-grained fields $\overline{f}$ and ${\Phi}$ neglecting
the internal energy of the fluctuations $\overline{\tilde f\tilde
\Phi}$. Therefore, the mass and the energy can be calculated at any time of the evolution from the coarse-grained field $\overline{f}$. By contrast, the moments $M_{n}$ with $n\ge 2$ will be called
{\it fragile integrals} because they are altered on the coarse-grained
scale since $\overline{f^{n}}\neq \overline{f}^{n}$ (the local moments
of the distribution are defined by $\overline{f^n}({\bf r},{\bf
v},t)=\int \rho \eta^{n}d\eta$). Therefore, only the moments of the
fine-grained field $M^{f.g.}_{n}=\overline{M_{n}[f]}=\int
\overline{f^{n}}d{\bf r}d{\bf v}$ are conserved, i.e.
\begin{equation}
\label{E6}
M^{f.g.}_{n}=\int \rho({\bf  r,v},\eta)\eta^{n} d{\bf r}d{\bf v}d\eta.
\end{equation}
The moments of the coarse-grained field $M^{c.g.}_{n}[\overline{f}]=\int
\overline{f}^{n}d{\bf r}d{\bf v}$ are not conserved
during the evolution since $M_{n}[\overline{f}]\neq
\overline{M_{n}[f]}$. In a sense, the moments $M^{f.g.}_{n}$ are
``hidden constraints'' because they are expressed in terms of the
fine-grained distribution $\rho({\bf r,v},\eta)$ and they cannot be
measured from the (observed) coarse-grained field. They can be only
computed from the initial conditions before the system has mixed or
from the fine-grained field. Since in many cases we do not know the
initial conditions nor the fine-grained field, they often appear as
``hidden''. Note that instead of conserving the fine-grained moments,
we can equivalently conserve the total hypervolume $\gamma(\eta)=\int
\rho d{\bf r}d{\bf v}$ of each level $\eta$.

After a complex evolution, we may expect the system to be in the most
probable, i.e. most mixed state, consistent with all the constraints
imposed by the dynamics. This, however, relies on a hypothesis of
ergodicity (efficient mixing) whose validity will be discussed in
Sec. 6. We define the mixing entropy as the logarithm of the number of
microscopic configurations associated with the same macroscopic state
characterized by the probability density $\rho({\bf r},{\bf
v},\eta)$. To get this number, we divide the macrocells $({\bf r},{\bf
r}+d{\bf r};{\bf v},{\bf v}+d{\bf v})$ into $\nu$ microcells of size
$h$ and denote by $n_{ij}$ the number of microcells occupied by the
level $\eta_{j}$ in the $i$-th macrocell. Note that a microcell can be
occupied only by one level $\eta_{j}$. This is due to the fact that we
make the statistical mechanics of a continuous field $f({\bf r},{\bf
v},t)$ instead of point particles. Therefore, we cannot ``compress''
that field, unlike point particles. A simple combinatorial analysis
indicates that the number of microstates associated with the
macrostate $\lbrace n_{ij}\rbrace$ is
\begin{equation}
W(\lbrace n_{ij}\rbrace)=\prod_{j}N_{j}!\prod_{i}{\nu!\over n_{ij}!},
\label{E7}
\end{equation}
where $N_{j}=\sum_{i}n_{ij}$ is the total number of microcells
occupied by $\eta_{j}$ (this is a conserved quantity equivalent to
$\gamma(\eta)$). We have to add the normalization condition
$\sum_{j}n_{ij}=\nu$, equivalent to Eq. (\ref{E1}), which prevents
overlapping of different levels (we note that we treat here the level
$\eta=0$ on the same footing as the others). This constraint plays
a role similar to the Pauli exclusion principle in quantum
mechanics. Morphologically, the Lynden-Bell statistics
(\ref{E7}) corresponds to a $4^{\rm th}$ type of statistics since
the particles are distinguishable but subject to an exclusion
principle \cite{lb}. There is no such exclusion for the
statistical equilibrium of point particles since they are free a
priori to approach each other, so we can put several particles in the
same microcell (see Sec. 2.4).

Taking the logarithm of $W$ and using the Stirling formula, we
get
\begin{equation}
\ln W(\lbrace n_{ij} \rbrace)=-\sum_{i,j}n_{ij}\ln
{n_{ij}}=-\sum_{i,j}\nu h^{2d}{\rho_{ij}}\ln {\rho_{ij}}, \label{E8}
\end{equation}
where $\rho_{ij}=\rho({\bf r}_{i},{\bf v}_{i},\eta_{j})=n_{ij}/\nu h^{2d}$.
Passing to the continuum limit
$\nu\rightarrow 0$, we obtain the  mixing
entropy introduced by Lynden-Bell \cite{lb}:
\begin{equation}
S_{LB}[\rho]=-\int \rho({\bf r},{\bf v},\eta)\ln\rho({\bf r},{\bf v},\eta)d{\bf r}d{\bf v}d\eta.
\label{E9}
\end{equation}
Note that the Lynden-Bell entropy can be interpreted as the Boltzmann
entropy for a distribution of levels $\eta$ (including
$\eta=0$). Equation (\ref{E9}) is sometimes called a {\it
collisionless entropy} to emphasize the distinction with the {\it
collisional entropy} (\ref{bol4}). Assuming ergodicity or ``efficient
mixing'', the statistical equilibrium state is obtained by maximizing
the Lynden-Bell entropy $S_{LB}[\rho]$ while conserving the mass $M$,
the energy $E$ and all the Casimirs (or moments $M_n$). We need also to
account for the local normalization condition (\ref{E1}).  This
problem is treated by introducing Lagrange multipliers, so that the
first order variations satisfy
\begin{equation}
\delta S_{LB}-\beta\delta E-\sum_{n\ge 1}\alpha_{n}\delta M_{n}-\int\zeta({\bf r},{\bf v})\delta\biggl (\int \rho({\bf r},{\bf v},\eta)d\eta\biggr )d{\bf r}d{\bf v}=0,
\label{E10}
\end{equation}
where ${ \beta}$ is the inverse temperature and ${\alpha }_{n}$ the
``chemical potential" associated with $M_n$. The resulting optimal
probability density is a Gibbs state which has the form
\begin{equation}
\label{E11} \rho({\bf r},{\bf v},\eta) ={1\over
Z}\chi(\eta)e^{-(\beta\epsilon+\alpha)\eta},
\end{equation}
where $\epsilon={v^2\over 2}+\Phi({\bf r})$ is the energy of a
particle by unit of mass. In writing Eq. (\ref{E11}), we have
distinguished the Lagrange multipliers $\alpha$ and $\beta$ associated
with the robust integrals $M$ and $E$ from the Lagrange multipliers
$\alpha_{n>1}$, associated with the conservation of the fragile
moments $M_{n>1}=\int \rho
\eta^{n}d\eta d{\bf r}d{\bf v}$, which have been regrouped in
the function $\chi(\eta)\equiv {\rm
exp}(-\sum_{n>1}\alpha_{n}\eta^{n})$. This distinction will make sense
in the following. Under this form, we see that the equilibrium
distribution of phase levels is a product of a universal Boltzmann
factor $e^{-(\beta\epsilon+\alpha)\eta}$ by a non-universal function
$\chi(\eta)$ depending on the initial conditions. The partition
function $Z$ is determined by the local normalization condition $\int
\rho d\eta=1$ leading to
\begin{equation}
\label{E12}
Z=\int \chi(\eta)e^{-\eta(\beta\epsilon+\alpha)}d\eta.
\end{equation}
Finally, the
equilibrium coarse-grained distribution function defined by $\overline{f}=\int
\rho \eta d\eta$ can be written
\begin{equation}
\label{E14}
\overline{f}={\int \chi(\eta)\eta e^{-\eta(\beta\epsilon+\alpha)}d\eta\over\int \chi(\eta) e^{-\eta(\beta\epsilon+\alpha)}d\eta},
\end{equation}
or, equivalently,
\begin{equation}
\label{E13}
\overline{f}=-{1\over\beta}{\partial\ln Z\over\partial \epsilon}=F(\beta\epsilon+\alpha)=\overline{f}(\epsilon).
\end{equation}
The function $F$ is given by
\begin{equation}
\label{g8} F(x)=-(\ln\hat\chi)'(x),
\end{equation}
where we have defined
$\hat{\chi}(x)=\int_{0}^{+\infty}\chi(\eta)e^{-\eta x}d\eta$.  It is
straightforward to check that a distribution function $f=f(\epsilon)$
depending only on the energy $\epsilon$ is a stationary solution of
the Vlasov equation \cite{bt}.  Therefore, for a given initial
condition, the statistical theory of Lynden-Bell selects a particular
stationary solution of the Vlasov equation (most mixed) among all
possible ones (an infinity!) \footnote{Incidentally, the fact that the
coarse-grained distribution function should be a stationary solution
of the Vlasov equation is not obvious. This depends on the definition
of coarse-graining \cite{cb}.}.  Specifically, the equilibrium state
is obtained by solving the integro-differential equation
\begin{equation}
\label{vp2c}
{\Phi}({\bf r})=\int u({\bf r}-{\bf r}')\overline{f}_{\alpha_{n},\beta} \lbrack {v^{'2}}/{2}+\Phi({\bf r}') \rbrack d{\bf r}'d{\bf v}',
\end{equation}
and relating the Lagrange multipliers $\alpha_{n}$, $\beta$ to the
constraints $M_{n}$, $E$. Then, we have to make sure that the
distribution is a maximum of $S_{LB}$, not a minimum or a saddle
point.  We note that the coarse-grained distribution function
$\overline{f}(\epsilon)$ predicted by Lynden-Bell can take a wide
diversity of forms depending on the function $\chi(\eta)$ determined
by the fragile moments (``hidden constraints'') fixed by the initial
condition.  The coarse-grained distribution function (\ref{E14}) can
be viewed as a sort of {\it superstatistics} as it is expressed as a
superposition of Boltzmann factors (on the fine-grained scale)
weighted by a non-universal function $\chi(\eta)$
\cite{super}. Some examples will be given in the sequel. In
the present context, the function $\chi(\eta)$ is determined from the
constraints {\it a posteriori}. Indeed, we have to solve the full
problem in order to get the expression of $\chi(\eta)$. In this sense,
the constraints associated with the conservation of the fine-grained
moments are treated microcanonically \footnote{In the context of
two-dimensional turbulence, some authors \cite{ellisgeo,aussois} have
proposed that the vorticity distribution $\chi(\sigma)$ could be
imposed by a small-scale forcing, in which case it should be treated
canonically (as a {\it prior} distribution). It is not clear whether
there exists a similar interpretation for other systems with
long-range interactions (note, however, that a small-scale forcing is
present in certain models of dark matter in astrophysics). Here, we
exclusively consider {\it isolated} systems where the distribution
$\chi(\eta)$ is determined implicitly by the initial conditions.}. We
emphasize that the function $\overline{f}(\epsilon)$ determining the
metaequilibrium state depends on the details of the initial
conditions. This differs from the ordinary statistical equilibrium
state (\ref{bol9}) which only depends on the mass $M$ and the energy
$E$. Here, we need to know the value of the fine-grained moments
$M_{n>1}^{f.g.}$ which are accessible only in the initial condition
(or from the fine-grained field) since the {\it observed} moments are
altered for $t>0$ by the coarse-graining as the system undergoes a
mixing process ($M_{n>1}^{c.g.}\neq M_{n>1}^{f.g.}$). This makes the
practical prediction of $\overline{f}(\epsilon)$ very complicated, or
even impossible, since, in practice, we often do not know the initial
conditions in detail (e.g., the initial state giving rise to a galaxy
or to a large scale vortex). In addition, in many cases, we cannot be
sure that the initial condition is not already mixed
(coarse-grained). If it has a fine-grained structure, this would
change a priori the prediction of the metaequilibrium state (see
p. 284 of \cite{JFM} and \cite{Arad}).

We note that the coarse-grained distribution function predicted by Lynden-Bell depends
only on the individual energy $\epsilon$ of the particles.  According
to the Jeans theorem \cite{bt}, such distribution functions form just
a particular class of stationary solutions of the Vlasov equation.
 We also note
that $\overline{f}(\epsilon)$ is a monotonically decreasing function
of energy.  Indeed,
from Eqs. (\ref{E11}) and (\ref{E13}), it is easy to establish that
\begin{equation}
\label{g1} \overline{f}'(\epsilon)=-\beta f_{2},\qquad
f_{2}\equiv \int \rho (\eta-\overline{f})^{2}d\eta\ge 0,
\end{equation}
where $f_{2}\equiv \overline{f^{2}}-\overline{f}^{2}$ is the centered
local variance of the distribution $\rho({\bf r},{\bf
v},\eta)$. Therefore, $\overline{f}'(\epsilon)\le 0$ since $\beta\ge
0$ is required to make the velocity profile normalizable. We also have
$F'(x)=-f_{2}(x)$ so that $F$ is a monotonically decreasing
function. Finally, the coarse-grained distribution function satisfies
$\overline{f}({\bf r},{\bf v})\le f_{0}^{max}$ where
$f_{0}^{max}=f_{max}({\bf r},{\bf v},t=0)$ is the maximum value of the
initial (fine-grained) distribution function. This inequality can be
obtained from Eq. (\ref{E14}) by taking the limit $\epsilon\rightarrow
-\infty$ for which $\overline{f}(\epsilon)\rightarrow
\eta_{max}=f_{0}^{max}$ and using the fact that $f(\epsilon)$ is a
decreasing function. Of course, the inequality $0\le \overline{f}\le
f_{0}^{max}$ is clear from physical considerations since the
coarse-grained distribution function locally averages over the
fine-grained levels. Since the fine-grained distribution function is
conserved by the Vlasov equation, the coarse-grained distribution
function is always intermediate between the minimum and the maximum
values of $f_{0}$.  We also note \cite{leprovost} that the most
probable level $\eta_{*}({\bf r},{\bf v})$ of the distribution
(\ref{E11}) is given by $
\eta_{*}=\lbrack (\ln\chi)'\rbrack^{-1}(\beta\epsilon+\alpha)$,
provided that $(\ln\chi)''(\eta_{*})<0$.  The function $\eta_{*}({\bf
r},{\bf v})$ is a stationary solution of the Vlasov equation of the
form $\eta_{*}=\eta_{*}(\epsilon)$ which monotonically decreases with the
energy since $\eta_{*}'(\epsilon)=\beta/(\ln\chi)''(\eta_{*})<0$.  It
usually differs from the average value $\overline{f}({\bf r},{\bf v})$
of the distribution (\ref{E11}).

If the initial DF takes only two values $f_0=0$ and
$f_0=\eta_{0}$, the Lynden-Bell entropy becomes
\begin{equation}
\label{sfd}
S_{LB}=-\int \biggl\lbrace \frac{\overline{f}}{\eta_{0}}\ln \frac{\overline{f}}{\eta_{0}}+\biggl (1- \frac{\overline{f}}{\eta_{0}}\biggr )\ln \biggl (1-\frac{\overline{f}}{\eta_{0}}\biggr )\biggr\rbrace d{\bf r}d{\bf v},
\end{equation}
which is similar to the Fermi-Dirac entropy (but with, of course, a
completely different interpretation). In that case, the statistical
prediction of Lynden-Bell for the metaequilibrium state resulting from
a ``collisionless'' violent relaxation is
\cite{lb,csr,csmnras}:
\begin{equation}
\label{E15}
\overline{f}={\eta_{0}\over 1+ e^{\eta_{0}(\beta\epsilon+\alpha)}},
\end{equation}
which is similar to the Fermi-Dirac distribution \footnote{Degeneracy
effects associated with the Lynden-Bell statistics (\ref{E15}) have
been observed in 2D turbulence \cite{staquet}, self-gravitating
systems \cite{hohl} and for the HMF model \cite{anto}.}.  This has to
be contrasted from the statistical equilibrium state resulting from a
``collisional'' relaxation which is the Maxwell-Boltzmann distribution
(\ref{bol9}). In the dilute limit of Lynden-Bell's theory
$\overline{f}\ll \eta_{0}$, the DF (\ref{E15}) becomes
\begin{equation}
\label{E17}
\overline{f}=A e^{-\beta\eta_{0}\epsilon}.
\end{equation}
This is similar to the  statistical equilibrium state
(\ref{bol9}). Therefore, in this approximation, collisional and
collisionless relaxation lead to the same distribution function (the
mean field Maxwell-Boltzmann distribution) but with a completely
different interpretation, corresponding to very different
timescales. To emphasize the difference, note in particular the bar on
$f$ in Eq. (\ref{E17}) and the fact that the mass of the individual
stars $m$ in Eq. (\ref{bol9}) is replaced by the value $\eta_{0}$ of the
fine-grained distribution function in Eq. (\ref{E17}).

In conclusion, assuming ergodicity, the Lynden-Bell statistical
equilibium state $\rho({\bf r},{\bf v},\eta)$ is obtained by solving
the maximization problem
\begin{eqnarray}
\label{mlbm}
\max_{\rho}\quad \lbrace S_{LB}[\rho]\quad |\quad E[\rho]=E, \ M[\rho]=M, \ M^{f.g.}_{n>1}[\rho]=M^{f.g.}_{n>1}, \int \rho d\eta=1 \rbrace.
\end{eqnarray}
This corresponds to a condition of microcanonical stability. Let us introduce the Lynden-Bell free energy
$F_{LB}[\rho]=E[\rho]-T S_{LB}[\rho]$. The condition of canonical stability is 
\begin{eqnarray}
\label{mlbc}
\min_{\rho}\quad \lbrace F_{LB}[\rho]\quad |\quad M[\rho]=M, \ M^{f.g.}_{n>1}[\rho]=M^{f.g.}_{n>1}, \int \rho d\eta=1 \rbrace.
\end{eqnarray}
The optimization problems (\ref{mlbm}) and (\ref{mlbc}) have the same critical points. Furthermore, if $\rho({\bf r},{\bf v},\eta)$
solves the minimization problem (\ref{mlbc}) then it solves the maximization problem (\ref{mlbm}) and represents
therefore a maximum entropy state. However, the reciprocal is wrong in
case of ensemble inequivalence. A solution of (\ref{mlbm})  is not
necessarily a solution of (\ref{mlbc}): the minimization problem
(\ref{mlbc}) just determines a subclass of the solutions of the
maximization problem (\ref{mlbm}).  Therefore, the condition of canonical
stability (\ref{mlbc}) provides just a {\it sufficient} condition of
microcanonical stability: (\ref{mlbc}) $\Rightarrow$ (\ref{mlbm}).

{\bf Numerical algorithms:} A relaxation equation has been proposed to
solve the maximization problem (\ref{mlbm}). It is obtained from a
Maximum Entropy Production Principle (MEPP) and has the form of a
generalized mean field Fokker-Planck equation \cite{csr}. It can serve
as a numerical algorithm to determine the Lynden-Bell maximum entropy
state for given $E$, $M$ and $M_{n>1}^{f.g.}$ specified by the initial
conditions. The relaxation equation solving the microcanonical problem
(\ref{mlbm}) is
\begin{equation}
{\partial\rho\over\partial t}+{\bf v}\cdot {\partial\rho\over\partial {\bf r}}-\nabla\Phi\cdot {\partial\rho\over\partial {\bf v}}={\partial\over\partial {\bf v}}\cdot \biggl\lbrace D\biggl\lbrack {\partial\rho\over\partial {\bf v}}+\beta(t)(\eta-\overline{f})\rho {\bf v}\biggr\rbrack\biggr\rbrace,
\label{rlbm}
\end{equation}
with
\begin{equation}
\beta(t)=-\frac{\int D\frac{\partial \overline{f}}{\partial {\bf v}}\cdot {\bf v} d{\bf r}d{\bf v}}{\int D f_{2}v^{2}d{\bf r}d{\bf v}}.
\label{blb}
\end{equation}
This equation conserves the mass $M$, the energy $E$ (through the time
dependent Lagrange multiplier $\beta(t)$) and the fine-grained moments
$M^{f.g.}_{n>1}$, and monotonically increases the Lynden-Bell entropy:
$\dot S_{LB}\ge 0$ ($H$-theorem). Furthermore, $\dot S_{LB}=0$ iff
$\rho$ is a steady solution of Eq. (\ref{rlbm}). Therefore,
$S_{LB}[\rho]$ is the Lyapunov functional of the relaxation equation
(\ref{rlbm}). It results that $\rho({\bf r},{\bf v},\eta)$ is a
linearly dynamically stable stationary solution of Eq. (\ref{rlbm})
iff it is a (local) maximum of $S_{LB}$ at fixed $E$, $M$ and
$M^{f.g.}_{n>1}$. Minima or saddle points of entropy are dynamically
unstable. Furthermore, if $S_{LB}$ is bounded from above (within the
previous constraints) we can conclude for Lyapunov's theory that
$\rho({\bf r},{\bf v},\eta,t)$ will relax towards a (local) maximum of
$S_{LB}$ at fixed $E$, $M$ and $M^{f.g.}_{n>1}$ for $t\rightarrow
+\infty$ (if there exists several local entropy maxima, the selection
will depend on a complicated notion of basin of attraction).
Therefore, Eq. (\ref{rlbm}) with Eq. (\ref{blb}) can serve as a
numerical algorithm to solve the microcanonical problem (\ref{mlbm})
and compute the Lynden-Bell statistical equilibrium state. It can also
provide a phenomenological description of the out-of-equilibrium
dynamics towards Lynden-Bell's statistical equilibrium \cite{csr}.

If we fix the inverse temperature $\beta$ instead of the energy in
Eq. (\ref{rlbm}), we get a relaxation equation solving the canonical
problem (\ref{mlbc}). This equation conserves the mass $M$ and the
fine-grained moments $M^{f.g.}_{n>1}$, and monotonically decreases the
Lynden-Bell free energy: $\dot F_{LB}\le 0$. Furthermore, $\dot
F_{LB}=0$ iff $\rho$ is a steady solution of (\ref{rlbm}) with fixed
$\beta$. Therefore, $F_{LB}[\rho]$ is the Lyapunov functional of this
relaxation equation. It results that $\rho({\bf r},{\bf v},\eta)$ is a
linearly dynamically stable stationary solution of Eq. (\ref{rlbm})
with fixed $\beta$ iff it is a (local) minimum of $F_{LB}$ at fixed
$M$ and $M^{f.g.}_{n>1}$. Furthermore, if $F_{LB}$ is bounded from
below, $\rho({\bf r},{\bf v},\eta,t)$ will relax towards a (local)
minimum of $F_{LB}$ at fixed $M$ and $M^{f.g.}_{n>1}$. Therefore,
Eq. (\ref{rlbm}) with fixed $\beta$ can serve as a numerical algorithm
to solve the canonical problem (\ref{mlbc}) and determine a subclass
of Lynden-Bell statistical equilibrium states since (\ref{mlbc})
$\Rightarrow$ (\ref{mlbm}).

\section{5. Generalized entropies in the reduced space of coarse-grained distributions}

Assuming ergodicity, we have seen that the most probable local
distribution of phase levels $\rho({\bf r},{\bf v},\eta)$ maximizes
the Lynden-Bell mixing entropy (\ref{E9}) while conserving mass,
energy and all the fine-grained moments. This functional of $\rho$ is
the proper form of Boltzmann entropy in the context of the violent
relaxation. It is obtained by a combinatorial analysis taking into
account the specificities of the collisionless (Vlasov) evolution. We
shall first show that the equilibrium coarse-grained distribution
function $\overline{f}({\bf r},{\bf v})$ (which is the function
directly accessible to the observations) extremizes a certain
functional $S[\overline{f}]$ at fixed mass $M$ and energy $E$
\cite{pre,super}. This functional is {\it non-universal} and depends on the
initial conditions. It is determined indirectly by the statistical
theory of Lynden-Bell and cannot be obtained from a combinatorial
analysis, unlike $S[\rho]$. Such functionals arise because they
encapsulate the influence of fine-grained constraints (Casimirs) that
are not accessible on the coarse-grained scale. They play the role of
``hidden constraints''. Now, if the distribution of phase levels
$\chi(\eta)$ is treated canonically, it can be shown that the
equilibrium coarse-grained distribution function $\overline{f}({\bf
r},{\bf v})$ maximizes $S[\overline{f}]$ at fixed mass $M$ and energy
$E$. In this sense, $S[\overline{f}]$ will be called a ``generalized
entropy'' in the reduced space of coarse-grained distributions
\footnote{We emphasize that this notion of ``generalized entropies'' 
\cite{pre,super} is completely different from the notion of generalized entropies introduced by Tsallis (see Sec. 7). These entropies take into account the influence of the Casimir constraints
for an ergodic evolution while the Tsallis entropies aim at describing
non-ergodic behaviours. }. A coarse-grained DF that maximizes a
``generalized entropy'' $S[\overline{f}]$ at fixed mass and energy is
a Lynden-Bell statistical equilibrium state (but the reciprocal is
wrong in case of ensemble inequivalence).  We note that the entropic
functionals $S[\rho]$ and $S[\overline{f}]$ are defined on two
different spaces. The $\rho$-space is the relevant one to make the
statistical mechanics of violent relaxation \cite{lb}. The
$\overline{f}$-space is a sort of projection of the $\rho$-space in
the space of directly observable (coarse-grained) distributions.

Since the coarse-grained distribution function
$\overline{f}(\epsilon)$ predicted by the statistical theory of
Lynden-Bell depends only on the individual energy and is monotonically
decreasing (see Sec. 4), it extremizes a functional of the form
\cite{pre,super}:
\begin{equation}
\label{g2} S[\overline{f}]=-\int C(\overline{f}) d{\bf r}d{\bf v},
\end{equation}
at fixed mass $M$ and energy $E$, where $C(\overline{f})$ is a convex function,
i.e. $C''>0$. Indeed, introducing Lagrange multipliers and writing the
variational principle as
\begin{equation}
\label{g3} \delta S-\beta \delta E-\alpha\delta M=0,
\end{equation}
we find that
\begin{equation}
\label{g4} C'(\overline{f})=-\beta\epsilon-\alpha.
\end{equation}
Since $C'$ is a monotonically increasing function of $\overline{f}$, we can
inverse this relation to obtain
\begin{equation}
\label{g5} \overline{f}=F(\beta\epsilon+\alpha)=\overline{f}(\epsilon),
\end{equation}
where
\begin{equation}
\label{g6} F(x)=(C')^{-1}(-x).
\end{equation}
From the identity
\begin{equation}
\label{g7} \overline{f}'(\epsilon)=-\beta/C''(\overline{f}),
\end{equation}
resulting from Eq. (\ref{g4}), $\overline{f}(\epsilon)$ is a
monotonically decreasing function of energy (if $\beta>0$). Thus,
Eq. (\ref{E13}) is compatible with Eq. (\ref{g5}) provided that we identify (\ref{g6}) with (\ref{g8}). Therefore, to any function $F(x)$
determined by the function $\chi(\eta)$ in the statistical theory
according to Eq. (\ref{g8}), we can associate to the metaequilibrium
state (\ref{E13}) a functional (\ref{g2}) where $C(\overline{f})$ is
given by Eq. (\ref{g6}) or equivalently by
\begin{equation}
\label{g9} C(\overline{f})=-\int^{\overline{f}}F^{-1}(x)dx.
\end{equation}
Using Eq. (\ref{g8}), this function is explicitly given in terms of $\hat{\chi}$ by
\begin{equation}
\label{g10} C(\overline{f})=-\int^{\overline{f}}\lbrack (\ln\hat\chi)'\rbrack^{-1}(-x)dx.
\end{equation}
On the other hand, comparing Eqs. (\ref{g1}) and (\ref{g7}), we find that
\begin{equation}
\label{g11}
f_{2}={1}/{C''(\overline{f})}.
\end{equation}
This equation relates the variance of the equilibrium distribution
$f_{2}$ to the coarse-grained distribution function $\overline{f}$
through the function $C$. We note that $C(\overline{f})$ is a {\it
non-universal} function which depends on the initial
conditions. Indeed, it is determined by the function $\chi(\eta)$
which depends indirectly on the initial conditions through the
complicated procedures discussed in Sec. 4. In general,
$S[\overline{f}]$ is {\it not} the Boltzmann functional
$S_B[\overline{f}]=-\int \overline{f}\ln \overline{f} d{\bf r}d{\bf
v}$ (except in the dilute limit of the Lynden-Bell theory) due to
fine-grained constraints (Casimirs) that modify the form of entropy
that we would naively expect. This is why the quasi-stationary state is
described by non-standard distributions (even for an assumed ergodic
evolution).  The existence of ``hidden constraints'' (here the Casimir
invariants that are not accessible on the coarse-grained scale) is the
physical reason for the occurrence of non-standard distributions and
``generalized entropies'' in our problem. {\it In fact, the
distribution is standard (Boltzmann) at the level of the local
distribution of fluctuations $\rho({\bf r},{\bf v},\eta)$
($\rho$-space) and non-standard at the level of the macroscopic
coarse-grained field $\overline{f}({\bf r},{\bf v})$
($\overline{f}$-space).}

So far, we have shown that the coarse-grained DF extremizes the
functional (\ref{g2})-(\ref{g10}) at fixed mass and energy. The
conditions under which it {\it maximizes} this functional at fixed
mass and energy have been discussed recently by Bouchet
\cite{baussois}. In the following, we give a complementary discussion, 
adopting a presentation similar to the one used in \cite{aaantonov} in
a different context (see also Sec. 11). We introduce the grand entropy \cite{super}:
\begin{equation}
\label{t4}
S_{\chi}[\rho]= S_{LB}[\rho]-\sum_{n>1}\alpha_{n}\ M^{f.g.}_{n}[\rho],
\end{equation}
which is the Legendre transform of $S_{LB}[\rho]$ with respect to the
fragile moments $M^{f.g.}_{n>1}[\rho]$. Expliciting these fine-grained
moments, it can be rewritten
\begin{equation}
\label{t5}
S_{\chi}\lbrack \rho\rbrack=-\int \rho\ \ln\biggl\lbrack {\rho\over \chi(\eta)}\biggr\rbrack \ d{\bf r}d{\bf v}d\eta,
\end{equation}
where we have noted
\begin{equation}
\label{t6}
\chi(\eta)\equiv{\rm exp}\biggl\lbrace -\sum_{n>1}\alpha_{n}\eta^{n}\biggr\rbrace.
\end{equation}
We now consider the grand microcanonical ensemble defined by \cite{super}:
\begin{eqnarray}
\label{mf1}
\max_{\rho}\quad \lbrace S_{\chi}[\rho]\quad |\quad E[\rho]=E, M[\rho]=M, \int\rho d\eta=1  \rbrace.
\end{eqnarray}
If $\rho({\bf r},{\bf v},\eta)$ solves the maximization problem
(\ref{mf1}), then it solves the maximization problem (\ref{mlbm}) and
represents therefore a Lynden-Bell maximum entropy state. Note that
the reciprocal is wrong in case of ensemble inequivalence
\cite{ellis,bb}: a solution of (\ref{mlbm})  is not necessarily a
solution of (\ref{mf1}). In other words, the maximization problem
(\ref{mf1}) determines only a sublclass of solutions of the
maximization problem (\ref{mlbm}). Therefore, it only provides a {\it
sufficient} condition of Lynden-Bell thermodynamical stability. Note
that we can also consider the grand canonical ensemble
\begin{eqnarray}
\label{mf1c}
\min_{\rho}\quad \lbrace F_{\chi}[\rho]=E-TS_{\chi}[\rho]\quad |\quad M[\rho]=M, \int\rho d\eta=1   \rbrace,
\end{eqnarray}
where $F_{\chi}[\rho]$ is a grand free energy.  We have the
implications: (\ref{mf1c}) $\Rightarrow$ (\ref{mf1}) $\Rightarrow$
(\ref{mlbm}) and (\ref{mf1c}) $\Rightarrow$ (\ref{mlbc}) $\Rightarrow$
(\ref{mlbm}) so that (\ref{mf1c}) gives a sufficient condition of
Lynden-Bell thermodynamical stability (less refined than
(\ref{mf1})). In the following, we shall consider the maximization
problem (\ref{mf1}). To determine the distribution $\rho_*({\bf
r},{\bf v},\eta)$ which maximizes $S_{\chi}[\rho]$ with the robust
constraints $E[\overline{f}]=E$, $M[\overline{f}]=M$, and the
normalization condition $\int\rho\, d\eta=1$, we can proceed in two
steps (we consider here the global maximization problem. The local
maximization problem is treated in the Appendix). {\it First step:} we
determine the distribution $\rho_1({\bf r},{\bf v},\eta)$ which
maximizes $S_{\chi}[\rho]$ with the constraint $\int\rho\, d\eta=1$
and a fixed coarse-grained distribution $\int\rho\eta \, d\eta
=\overline{f}({\bf r},{\bf v})$ (note that fixing the coarse-grained
distribution function automatically determines the robust constraints
$E$ and $M$). This gives a distribution $\rho_1[\overline{f}({\bf
r},{\bf v}),\eta]$ depending on $\overline{f}({\bf r},{\bf v})$ and
$\eta$. Substituting this distribution in the functional
$S_{\chi}[\rho]$, we obtain a functional $S[\overline{f}]\equiv
S_{\chi}[\rho_1]$ of the coarse-grained distribution $\overline{f}$.
{\it Second step:} we determine the distribution function
$\overline{f}_*({\bf r},{\bf v})$ which maximizes $S[\overline{f}]$
with the constraints $E[\overline{f}]=E$ and
$M[\overline{f}]=M$. Finally, we have $\rho_*({\bf r},{\bf
v},\eta)=\rho_1[\overline{f}_*({\bf r},{\bf v}),\eta]$. Let us now
show that $S[\overline{f}]\equiv S_{\chi}[\rho_1]$ is the functional
(\ref{g10}).  The distribution $\rho_1({\bf r},{\bf v},\eta)$ that
extremizes $S_{\chi}[\rho]$ with the constraints $\int\rho\, d\eta=1$
and $\int\rho\eta \, d\eta =\overline{f}({\bf r},{\bf v})$ satisfies
the first order variations
\begin{eqnarray}
\label{js1}
\delta S_{\chi}-\int \Phi({\bf r},{\bf v}) \delta \left  ( \int \rho
\eta d\eta\right ) d{\bf r}d{\bf v}-\int \zeta({\bf r},{\bf v}) \delta \left (\int \rho d\eta\right )
d{\bf r}d{\bf v}=0,
\end{eqnarray}
where $\Phi({\bf r},{\bf v})$ and $\zeta({\bf
r},{\bf v})$ are Lagrange multipliers.  This yields
\begin{eqnarray}
\rho_1({\bf r},{\bf v},\eta)=\frac{1}{Z({\bf
r},{\bf v})}\chi(\eta)e^{-\eta\Phi({\bf r},{\bf v})},\label{jsq}
\end{eqnarray}
where $Z({\bf r},{\bf v})$ and $\Phi({\bf r},{\bf v})$ are determined by
\begin{eqnarray} Z({\bf r},{\bf v})=\int \chi(\eta)e^{-\eta\Phi({\bf
r},{\bf v})}d\eta= \hat{\chi}(\Phi),\label{js2}
\end{eqnarray}
\begin{eqnarray}
\overline{f}({\bf r},{\bf v})=\frac{1}{Z({\bf r},{\bf v})}\int \chi(\eta)\eta
e^{-\eta\Phi({\bf r},{\bf v})}d\eta=-(\ln\hat{\chi})'(\Phi).\label{js3}
\end{eqnarray}
The critical point (\ref{jsq}) is a {\it maximum} of $S_{\chi}[\rho]$ with the
above-mentioned constraints since $\delta^{2} S_{\chi}=-\int
\lbrack (\delta\rho)^{2}/\rho\rbrack d{\bf r}d{\bf v}d\eta\le 0$. Then,
we compute
\begin{eqnarray}
S_{\chi}[\rho_1]=-\int \rho_1(-\eta\Phi-\ln\hat\chi )\, d{\bf
r}d{\bf v}d\eta=\int (\overline{f}\Phi+\ln\hat{\chi}(\Phi))  \, d{\bf
r}d{\bf v}.\nonumber\\
\label{js4}
\end{eqnarray}
Therefore, the functional of the coarse-grained DF is
\begin{eqnarray}
S[\overline{f}]=-\int C(\overline{f}) \, d{\bf r}d{\bf v},\label{fo}
\end{eqnarray}
with
\begin{eqnarray}
C(\overline{f})= -\overline{f}\Phi-\ln\hat{\chi}(\Phi).\label{js5}
\end{eqnarray}
Now $\Phi({\bf r},{\bf v})$ is related to $\overline{f}(\bf{r},{\bf v})$ by
Eq. (\ref{js3}). This implies that
\begin{eqnarray}
C'(\overline{f})=-
\Phi=-[(\ln\hat\chi)']^{-1}(-\overline{f}),\label{js6}
\end{eqnarray}
so that
\begin{eqnarray}
C(\overline{f})=-\int^{\overline{f}}
[(\ln\hat\chi)']^{-1}(-x)dx.\label{rel}
\end{eqnarray}
This returns the functional (\ref{g10}). We now conclude that
$\rho({\bf r},{\bf v},\eta)$ solves the maximization problem
(\ref{mf1}) if, and only if, $\overline{f}({\bf r},{\bf v})$ solves
the maximization problem
\begin{eqnarray}
\label{projm}
\max_{\overline{f}}\quad \lbrace S[\overline{f}]\quad |\quad E[\overline{f}]=E, M[\overline{f}]=M  \rbrace,
\end{eqnarray}
where $S[\overline{f}]$ is given by Eqs. (\ref{fo}) and (\ref{rel}).
The optimal distributions are related to each other by $\rho_{*}({\bf
r},{\bf v},\eta)=\rho_1[\overline{f}_*({\bf r},{\bf v}),\eta]$
according to the procedure described previously. We have the
implications: (\ref{projm}) $\Leftrightarrow$ (\ref{mf1})
$\Rightarrow$ (\ref{mlbm}). Therefore, if $\overline{f}({\bf r},{\bf
v})$ solves the maximization problem (\ref{projm}), the corresponding
distribution $\rho({\bf r},{\bf v},\eta)$ solves the maximization
problem (\ref{mlbm}) and is therefore a Lynden-Bell statistical
equilibrium state. However, the reciprocal is wrong in case of
ensemble inequivalence. The maximum entropy state satisfying
(\ref{mlbm}) may not satisfy (\ref{mf1}) or (\ref{projm}). Therefore,
although the coarse-grained DF always extremizes $S[\overline{f}]$ at
fixed $E$, $M$, it is {\it not} a maximum of $S[\overline{f}]$ at
fixed $E$, $M$ in the region of ensemble inequivalence.  These
maximization problems (\ref{mf1}) or (\ref{projm}) provide only {\it
sufficient} conditions of Lynden-Bell thermodynamical stability
\footnote{Note that the maximization problem (\ref{projm}) determines
a state that is both Lynden-Bell thermodynamically stable with respect to fine
grained perturbations $\delta\rho({\bf r},{\bf v},\eta)$ (according to
criterion (\ref{mlbm}) $\Leftarrow$ (\ref{projm})) and formally nonlinearly
dynamically Vlasov stable with respect to coarse-grained perturbations
$\delta\overline{f}({\bf r},{\bf v})$ (according to criterion
(\ref{dynm})). On the other hand, since (\ref{mlbm}) does not imply
(\ref{projm}) in case of ensemble inequivalence, we conclude that
{\it there may exist stable Lynden-Bell statistical equilibrium states that
do not satisfy the criterion of formal nonlinear dynamical stability
(\ref{dynm})}. Therefore, either: (i) the ensembles are always
equivalent with respect to the conjugate variables
$(M_{n>1}^{f.g.},\alpha_{n>1})$ (ii) Lynden-Bell thermodynamical
stability does not imply formal nonlinear dynamical stability (iii)
there exists a stronger condition of nonlinear dynamical stability
than (\ref{dynm}). This intriguing point needs further development. In
particular, it would be interesting to exhibit a situation of ensemble
inequivalence with respect to the conjugate variables
$(M_{n>1}^{f.g.},\alpha_{n>1})$. }. Returning to the maximization
problem (\ref{projm}), we introduce the generalized free energy
$F[\overline{f}]=E[\overline{f}]-T S[\overline{f}]$. Now, if
$\overline{f}({\bf r},{\bf v})$ solves the minimization problem
\begin{eqnarray}
\label{projc}
\min_{\overline{f}}\quad \lbrace F[\overline{f}]\quad |\quad M[\overline{f}]=M \rbrace,
\end{eqnarray}
then it solves the maximization problem (\ref{projm}). However, the
reciprocal is wrong in case of ensemble inequivalence: the solutions
of (\ref{projm})  are not necessarily solutions of
(\ref{projc}). Therefore, the generalized canonical problem
(\ref{projc}) determines only a sublclass of solutions of the
generalized microcanonical problem (\ref{projm}). We have the
implications: (\ref{projc}) $\Rightarrow$ (\ref{projm})
$\Leftrightarrow$ (\ref{mf1}) $\Rightarrow$ (\ref{mlbm}) and
(\ref{projc}) $\Leftrightarrow$ (\ref{mf1c}) $\Rightarrow$ (\ref{mf1})
$\Rightarrow$ (\ref{mlbm}) .

{\bf Numerical algorithms:} The numerical algorithms solving the
optimization problems (\ref{projm}) and (\ref{projc}) will be
presented in Sec. 10 in connection with the formal nonlinear dynamical
stability with respect to the Vlasov equation.

\section{6. Incomplete violent relaxation}

The statistical approach presented previously rests on the assumption
that the collisionless mixing is efficient so that the ergodic
hypothesis which sustains the statistical theory of Lynden-Bell is
fulfilled. There are situations where the Lynden-Bell prediction works
relatively well. However, there are other situations where the
Lynden-Bell prediction fails. It has been understood since the
beginning \cite{lb} that violent relaxation may be {\it incomplete} in
certain cases so that the Lynden-Bell mixing entropy is {\it not}
maximized in the whole available phase space. Incomplete relaxation
\cite{next05} can lead to more or less severe deviations from the
Lynden-Bell prediction. Physically, the system tends to reach the
Lynden-Bell maximum entropy state during violent relaxation but, in
some cases, it cannot attain it because the variations of the
potential, that are the engine of the evolution, die away before the
relaxation process is complete (there may be other reasons for
incomplete relaxation).  Since the Vlasov equation admits an infinite
number of stationary solutions, the coarse-grained distribution
$\overline{f}({\bf r},{\bf v},t)$ can be trapped in one of them and
remain frozen in that quasi stationary state $\overline{f}_{QSS}({\bf
r},{\bf v})$ until collisional effects finally come into play (on
longer timescales). This steady solution is not always the most mixed
state (it can be only partially mixed) so it may differ from
Lynden-Bell's statistical prediction. Thus, for dynamical reasons, the
system does not always explore the whole phase space ergodically.  In
general, the statistical theory of Lynden-Bell gives a relatively good
first order prediction of the QSS without fitting parameter and is
able to explain out-of-equilibrium phase transitions between different
types of structures, depending on the values of the control parameters
fixed by the initial condition. However, there are cases where the
prediction does not work well (it can sometimes be very bad) because
of incomplete relaxation. Other stationary solutions of the Vlasov
equation, differing from Lynden-Bell's prediction, can arise in case
of incomplete violent relaxation. The difficulty is that we do not
know a priori whether the prediction of Lynden-Bell will work or fail
because this depends on the dynamics and it is difficult to know in
advance if the system will mix well or not.  Let us give some examples
of complete and incomplete violent relaxation in stellar systems, 2D
turbulence and for the HMF model (for more discussion, see
\cite{next05,paper3}).

{\bf Stellar systems:} The concept of violent relaxation was first
introduced by Lynden-Bell
\cite{lb} to explain the apparent regularity of elliptical galaxies in
astrophysics. However, for 3D stellar systems the prediction of
Lynden-Bell leads to density profiles whose mass is infinite (the
density decreases as $r^{-2}$ at large distances). In other words,
there is no maximum entropy state at fixed mass and energy in an
unbounded domain. Furthermore, it is known that the distribution
functions (DF) of galaxies do not only depend on the energy
$\epsilon=v^{2}/2+\Phi({\bf r})$ contrary to what is predicted by the
Lynden-Bell statistical theory. This means that other ingredients are
necessary to understand their structure. However, the approach of
Lynden-Bell is able to explain why elliptical galaxies have an almost
isothermal core. Indeed, it is able to justify a Boltzmannian
distribution $\overline{f}\sim e^{-\beta\epsilon}$ in the core without
recourse to collisions which operate on a much longer timescale $
t_{R}\sim (N/\ln N)t_{D}$.  However, violent
relaxation is incomplete in the halo. The concept of incomplete
violent relaxation explains why galaxies are more confined than
predicted by statistical mechanics (the density profile of elliptical
galaxies decreases as $r^{-4}$ instead of $r^{-2}$ \cite{bt}).

For one dimensional self-gravitating systems, the Lynden-Bell entropy
has a global maximum at fixed mass and energy in an unbounded
domain. Early simulations of the 1D Vlasov-Poisson system starting
from a water-bag initial condition have shown a relatively good
agreement with the Lynden-Bell prediction \cite{hohl}. In other cases,
the Vlasov equation (and the corresponding $N$-body system) can have a very
complicated, non-ergodic, dynamics.  For example, starting from an
annulus in phase space, Mineau {\it et al.} \cite{mineau} have
observed the formation of phase-space holes which block the relaxation
towards the Lynden-Bell distribution. In that case, the system does
not even relax towards a stationary state of the Vlasov equation but
develops everlasting oscillations.

{\bf Two-dimensional vortices:} In the context of two-dimensional
turbulence, Miller \cite{miller} and Robert \& Sommeria \cite{rs} have
developed a statistical mechanics of the 2D Euler equation which is
similar to the Lynden-Bell theory (see
\cite{csr} for a description of this analogy). This theory works
relatively well to describe vortex merging \cite{rsmepp} or the
nonlinear development of the Kelvin-Helmholtz instability in a shear
layer \cite{staquet}. It can account for the numerous bifurcations
observed between different types of vortices (monopoles, dipoles,
tripoles,...) \cite{JFM,JFM2} and is able to reproduce the structure of
geophysical and jovian vortices like Jupiter's great red spot
\cite{bs,turk,aussois}.

However, some cases of incomplete relaxation have been reported.  For
example, in the plasma experiment of Huang
\& Driscoll \cite{hd}, the Miller-Robert-Sommeria (MRS) statistical theory
gives a reasonable prediction of the QSS without fit but the agreement
is not perfect \cite{brands}. The observed central density is larger
than predicted by theory and the tail decreases more rapidly than
predicted by theory, i.e. the vortex is more confined. This is related
to the fact that mixing is not very efficient in the core and in the
tail of the distribution (this can be explained by invoking the
concept of ``maximum entropy bubbles'' to delimitate the region where
the vorticity mixes well \cite{JFM2} or by developing a kinetic theory
of violent relaxation, see \cite{bbgky} and Sec. 8). As noted by
Boghosian \cite{boghosian}, in this case the QSS can be fitted by a
Tsallis distribution where the density drops to zero at a finite
distance.

{\bf The HMF model:} Recently, Antoniazzi {\it et al.} \cite{anto}
have performed numerical simulations of the HMF model, starting from a
water bag initial condition, to test the prediction of the Lynden-Bell
theory. For a given value of the energy $U=0.69$, this theory predicts
an out-of-equilibrium phase transition between a spatially homogeneous
phase and a spatially inhomogeneous phase above a critical
magnetization $M_{crit}=0.897$ discovered in
\cite{epjb,anto}. Numerical simulations show a relatively good
agreement with the Lynden-Bell theory when the initial magnetization
$M_{0}<M_{crit}$. However, for $M_{0}>M_{crit}$ several authors report
important deviations to the Lynden-Bell theory. In particular, for
$M_{0}=1$, Latora {\it et al.} \cite{latora} find that the QSS is
spatially homogeneous ($M_{QSS}=0$) and the velocity distribution
is non-gaussian, while the Lynden-Bell theory predicts a spatially
inhomogeneous state with gaussian tails (the {\it same} as the
Boltzmann equilibrium state, since we are in the non-degenerate limit
\cite{paper3}). Using isotropic water-bag initial conditions, Campa
{\it et al.} \cite{campa} show that the QSS is well-fitted by a
semi-elliptical distribution. As remarked in
\cite{paper3}, this corresponds to a Tsallis distribution with a
compact support (with index $q=3$) so that the distribution function
drops to zero above a finite velocity. This confinement is a typical
result of incomplete relaxation and can be explained by a kinetic
theory of violent relaxation (see \cite{paper4} and Sec. 8).

On the other hand, using different initial conditions, Morita \&
Kaneko \cite{mk} find that the system does not relax to a QSS but
exhibits oscillations whose duration diverges with $N$. Therefore, the
Lynden-Bell prediction clearly fails. This long-lasting periodic or
quasi periodic collective motion appears through Hopf bifurcation and
is due to the presence of clumps (high density regions) in phase
space. We remark that this behaviour is relatively similar to the one
reported by Mineau {\it et al.} \cite{mineau} for self-gravitating
systems, except that they observe phase space {\it holes} instead of
phase space {\it clumps}.

\section{7. Generalized thermodynamics: the Tsallis entropy}

To account for the non-Boltzmannian nature of the QSS, some authors
\cite{boghosian,latora,ts} have proposed to change the form of the
entropy. In particular, it has been proposed that the QSS could be in
the ``most probable'' state that maximizes the Tsallis entropy
\begin{equation}
S_{q}[f]=-{1\over q-1}\int (f^{q}({\bf r},{\bf
v})-f({\bf r},{\bf v}))d{\bf r}d{\bf v},
\label{se6}
\end{equation}
at fixed mass and energy. This leads to a distribution function of the
form ${f}_{q}({\bf r},{\bf v})=\lbrack
\mu-\beta(q-1)\epsilon/q\rbrack^{1/(q-1)}$ (where $\mu$ and $\beta$ are Lagrange multipliers determined by the mass and the energy) which coincides with the
Boltzmann distribution for $q=1$ and differs from it when $q\neq
1$. Therefore, the Tsallis entropy (\ref{se6}) can be viewed as a
generalization of the Boltzmann entropy \cite{tsallis}.  However, it
was remarked by Brands {\it et al.} \cite{brands} that the entropy
$S_{q}[f]$ can lead to inconsistencies when applied to the context of
the violent relaxation (in order to describe the QSS) where the
system's dynamics is governed by the Vlasov equation. Indeed,
the Tsallis distribution ${f}_{q}({\bf r},{\bf v})$ may not respect
the properties of the Vlasov equation, in particular the fact that the
DF is bounded by the maximum value of the initial condition: $f({\bf
r},{\bf v})\le f_{max}({\bf r},{\bf v},t=0)$. In our point of view
\cite{brands,super}, the proper form of Tsallis entropy in the context
of the violent relaxation is
\begin{equation}
S_{q}[\rho]=-{1\over q-1}\int (\rho^{q}({\bf r},{\bf
v},\eta)-\rho({\bf r},{\bf v},\eta))d{\bf r}d{\bf v}d\eta.
\label{se6b}
\end{equation}
This is a functional of the probability density $\rho({\bf r},{\bf
v},\eta)$ which takes into account the specificities of the
collisionless (Vlasov) dynamics. For $q\rightarrow 1$, it returns the
Lynden-Bell entropy (\ref{E9}). For $q\neq 1$, it could take into
account incomplete mixing and non-ergodicity. In that context, the $q$
parameter could be interpreted as a {\it measure of mixing} ($q=1$ if
the system mixes efficiently) and Tsallis entropy (\ref{se6b}) could
be interpreted as a functional attempting to take into account
non-ergodicity in the process of incomplete violent relaxation
\footnote{A striking property of the Tsallis entropy is to yield
distribution functions that have a compact support (for $q>1$) so that
the density vanishes above a maximum energy. This leads to a
confinement in phase space that is qualitatively (and sometimes
quantitatively) similar to what is observed in situations of
incomplete relaxation (see discussion in Sec. 6). Indeed, since mixing
is never complete, the high energy tail of the distribution is less
populated than what is predicted by the Boltzmann statistical
mechanics (note that this notion of confinement is similar to the
notion of maximum entropy ``bubble'' introduced in
\cite{JFM2}). However, as discussed in
\cite{next05}, it is not clear why this confinement should 
necessarily (universally) be described by a Tsallis
$q$-distribution. Other distributions with a compact support could
also result from an incomplete relaxation. }. Maximizing $S_{q}[\rho]$
at fixed mass, energy and Casimirs leads to a $q$-generalization of
the equilibrium state (\ref{E11}). This maximization principle is a
condition of thermodynamical stability (in Tsallis' generalized sense)
in the context of violent relaxation. Then, we can obtain a
$q$-generalization of the equilibrium coarse-grained distribution
function (\ref{E13}). In the case of two levels $f\in
\lbrace 0,\eta_{0}\rbrace$, and in the dilute limit of the theory
$\overline{f}\ll\eta_0$, $S_{q}[\rho]$ can be written in terms of the
coarse-grained distribution $\overline{f}=\rho\eta_{0}$ in the form
\begin{equation}
S_{q}[\overline{f}]=-{1\over q-1}\int \lbrack
(\overline{f}/\eta_0)^{q}-(\overline{f}/\eta_0)\rbrack d{\bf r}d{\bf
v},
\label{se6bb}
\end{equation}
similar to Eq. (\ref{se6}), but with a different interpretation than
the one given in \cite{boghosian,latora,ts}.  However, as stated in
\cite{brands}, it is not clear why complicated effects of
non-ergodicity (incomplete mixing) could be encapsulated in a simple
functional such as (\ref{se6b}). Other functionals of the form
$S=-\int C(\rho) d{\bf r}d{\bf v}d\eta$ (where $C$ is a convex
function) could be considered as well \footnote{We stress, however,
that the Tsallis entropy is the {\it simplest} generalization of the
Boltzmann entropy because, on an axiomatic point of view, the Tsallis
entropy possesses almost all the properties of the Boltzmann
entropy. This is not the case for other types of ``generalized
entropies''.}. Tsallis generalized entropy $S_{q}$ and Tsallis $q$
distributions can probably describe a certain type of non-ergodic
behaviour but not all of them: they are not ``universal attractors''
of the (coarse-grained) Vlasov equation.  For example, elliptical
galaxies are not stellar polytropes \cite{bt} so their DF cannot be
fitted by the Tsallis distribution
\cite{aa3,next05}.  An improved model is a {\it composite model} that
is isothermal in the core (justified by Lynden-Bell's theory of
violent relaxation) and polytropic in the halo (due to incomplete
relaxation) with an index $n=4$ corresponding to $q=7/5$
\cite{hm,aa3}. This observation suggests that the achieved
distributions could be described by entropies of the form (\ref{se6b})
where $q=q({\bf r},{\bf v})$ {\it depends on the position (in phase
space) so as to take into account the degree of mixing}
\cite{aa3}. Unfortunately, this idea is not very useful in practice if
we do not give an a priori prescription to relate the value of the
index $q({\bf r},{\bf v})$ to the phase-space region. Finally, we
note that the Tsallis distribution with a time dependent index $q(t)$
can provide a good fit of the distribution of the system during its
slow collisional evolution towards the Boltzmann distribution
($q=1$). This idea has been developed by Taruya \& Sakagami
\cite{tsprl} in stellar dynamics and proposed by Chavanis \cite{paper3} 
for the HMF model.

\section{8. Kinetic theory of violent relaxation}

An alternative idea to understand the problem of incomplete violent
relaxation is to develop a kinetic theory of the Vlasov equation on
the coarse-grained scale in order to understand dynamically what
limits the convergence towards the Lynden-Bell distribution \cite{csr}. Kinetic theories of collisionless relaxation have been developed in 
\cite{kp,sl,csr,quasivlasov,cb,paper4} with the aim to determine
the dynamical equation satisfied by the coarse-grained DF
$\overline{f}({\bf r},{\bf v},t)$. An interesting approach is based on
a quasilinear theory of the Vlasov equation
\cite{kp,sl,quasivlasov}.  In the two levels approximation, 
the quasilinear theory leads to a kinetic
equation for the coarse-grained DF of the form \cite{paper4}:
\begin{eqnarray}
\label{v10} {\partial \overline{f}\over\partial
t}+{\bf v}{\partial \overline{f}\over\partial
{\bf r}}-\nabla\overline{\Phi} {\partial \overline{f}\over \partial
{\bf v}}={\epsilon_{r}^{d}\epsilon_{v}^{d}}
{\partial \over
\partial v^{\mu}}\int_{0}^{t} d\tau\int d{\bf r}_{1}d{\bf v}_{1}
\frac{F^{\mu}}{m}(1\rightarrow 0) G(t,t-\tau) \nonumber\\
\times\frac{F^{\nu}}{m}(1\rightarrow 0)\biggl\lbrace  \overline{f}_1(\eta_{0}-\overline{f}_1){\partial
\overline{f}\over \partial v^{\nu}} -
\overline{f}(\eta_{0}-\overline{f}) {\partial \overline{f}_1\over
\partial v_1^{\nu}} \biggr \rbrace_{t-\tau},
\end{eqnarray}
where $\epsilon_{r}$, $\epsilon_{v}$ are the coarse-graining mesh
sizes in position and velocity spaces, $\overline{\Phi}$ is the
smooth field produced by the coarse-grained DF and $f=f({\bf r},{\bf v},t)$, $f_{1}=f({\bf r}_1,{\bf v}_1,t)$.  This equation is
expected to describe the late quiescent stages of the relaxation
process when the fluctuations have weaken so that the quasilinear
approximation can be implemented. It does not describe the early, very
chaotic, process of violent relaxation driven by the strong
fluctuations of the potential. The quasilinear theory of the
Vlasov equation is therefore a theory of ``quiescent'' collisionless
relaxation.

Equation (\ref{v10}) is very similar, in structure, to
Eq. (\ref{general}) for the collisional evolution, with nevertheless
three important differences: (i) the fluctuating force ${\cal
F}(1\rightarrow 0)$ is replaced by the direct force ${F}(1\rightarrow
0)$ because the fluctuations have a different nature in the two
problems. (ii) The distribution function $f$ in the collisional term
of Eq.  (\ref{general}) is replaced by the product
$\overline{f}(\eta_0-\overline{f})$ in Eq. (\ref{v10}). This nonlinear
term arises from the effective ``exclusion principle'', discovered by
Lynden-Bell, accounting for the non-overlapping of
phase levels in the collisionless regime. This is consistent with the
Fermi-Dirac-like entropy (\ref{sfd}) and Fermi-Dirac-like distribution
(\ref{E15}) at statistical equilibrium (iii) Considering the dilute
limit $\overline{f}\ll\eta_{0}$ to fix the ideas, we see that the
equations (\ref{v10}) and (\ref{general}) have the same mathematical
form differing only in the prefactors: the mass $m$ of a particle in
Eq. (\ref{general}) is replaced by the mass
$\eta_{0}\epsilon_r^{d}\epsilon_v^d$ of a completely filled macrocell
in Eq. (\ref{v10}). This implies that the timescales of collisional
and collisionless ``relaxation'' are in the ratio
\begin{eqnarray}
\frac{t_{ncoll}}{t_{coll}}\sim
\frac{m}{\eta_{0}\epsilon_r^{d}\epsilon_v^d}. \label{v12}
\end{eqnarray}
Since $\eta_{0}\epsilon_{r}^{d}\epsilon_{v}^{d}\gg m$, this ratio is
in general quite small implying that the collisionless relaxation is
much more rapid than the collisional relaxation. Typically,
$t_{ncoll}$ is of the order of a few dynamical times $t_{D}$ (its
precise value depends on the size of the mesh) while $t_{coll}$ is of
order $\sim {N}t_{D}$ or larger. The kinetic equation (\ref{v10})
conserves the mass and, presumably, the energy. By contrast, we cannot
prove an $H$-theorem for the Lynden-Bell entropy (\ref{sfd}). Indeed,
the time variation of the Lynden-Bell entropy is of the form
\begin{eqnarray}
\dot S_{LB}=\frac{1}{2}\epsilon_{r}^{d}\epsilon_{v}^{d}\int d{\bf r}d{\bf v}d{\bf r}_{1}d{\bf v}_{1}\frac{1}{\overline{f}(\eta_{0}-\overline{f})\overline{f}_{1}(\eta_{0}-\overline{f}_{1})}\int_{0}^{t}d\tau Q(t) G(t,t-\tau)Q(t-\tau),\label{g6fgh}
\end{eqnarray}
\begin{eqnarray}
Q(t)=\frac{{F}^{\mu}}{m}(1\rightarrow 0,t) \left\lbrack \overline{f}_{1}(\eta_{0}-\overline{f}_{1})\frac{\partial \overline{f}}{\partial v^{\mu}}-\overline{f}(\eta_{0}-\overline{f}) \frac{\partial \overline{f}_{1}}{\partial v_{1}^{\mu}}\right\rbrack,\label{g6fghb}
\end{eqnarray}
and its sign is not necessarily positive.  This depends on the
importance of memory effects. Now, even if Eq. (\ref{v10}) conserves the
energy and the mass and increases the Lynden-Bell entropy (\ref{sfd})
monotonically, this does not necessarily imply that the system will
converge towards the Lynden-Bell distribution (\ref{E15}). Indeed, it
has been observed in several experiments and numerical simulations
that the QSS does not exactly coincide with the strict statistical
equilibrium state predicted by Lynden-Bell because of the complicated
problem of {\it incomplete relaxation} (see Sec. 6). This is usually
explained by a lack of ergodicity or ``incomplete mixing''. Here, we
try to be a little more precise by using the kinetic theory. There can
be several reasons of incomplete relaxation:

(i) {\it Absence of resonances:} Very few is known concerning kinetic
equations of the form of Eq. (\ref{v10}) and it is not clear if the
Lynden-Bell distribution (\ref{E15}) is a stationary solution of that
equation (and if it is the only one). As explained in Sec. 3.4 for
the kinetic equation (\ref{general}) describing the collisional
relaxation, the relaxation may stop because the current ${\bf J}$
vanishes due to the {\it absence of resonances}. This argument may
also apply to Eq. (\ref{v10}) which has a similar structure and can be
a cause for incomplete relaxation. The system tries to approach the
statistical equilibrium state (as indicated by the increase of the
entropy) but may be trapped in a QSS that is different from the
statistical prediction (\ref{E15}). This QSS is a steady solution of
Eq.  (\ref{v10}) which cancels
individually the advective term (l.h.s.) and the effective collision
term (r.h.s.).  This determines a subclass of steady states of the
Vlasov equation (cancellation of the l.h.s.)  such that the
complicated ``turbulent'' current ${\bf J}$ in the r.h.s. vanishes.
This offers a large class of possible steady state solutions that can
explain the deviation between the QSS and the Lynden-Bell statistical
equilibrium state (\ref{E15}) observed, in certain cases, in
simulations and experiments of violent relaxation.

(ii) {\it Incomplete relaxation in phase space:} The turbulent current
${\bf J}$ in Eq. (\ref{v10}) is driven by the fluctuations
$f_{2}\equiv
\overline{\tilde{f}^{2}}=\overline{f^{2}}-\overline{f}^{2}$ of the
distribution function generating the fluctuations $\delta\Phi$ of the
potential (see \cite{paper4} for more details). In the ``mixing
region'' of phase space where the fluctuations are strong, the DF
tends to reach the Lynden-Bell distribution (\ref{E15}). As we depart
from the ``mixing region'', the fluctuations decay ($f_{2}\rightarrow
0$) and the mixing is less and less efficient $\|{\bf J}\|\rightarrow
0$. In these regions, the system takes a long time to reach the
Lynden-Bell distribution (\ref{E15}) and, in practice, cannot attain it
in the time available (see (iii)). In the two levels case, we have
$f_{2}=\overline{f}(\eta_{0}-\overline{f})$. Therefore, the phase
space regions where $\overline{f}\rightarrow 0$ or
$\overline{f}\rightarrow \eta_{0}$ do not mix well (the diffusion
current ${\bf J}$ is weak) and the observed DF can be sensibly
different from the Lynden-Bell distribution in these regions of phase
space. This concerns essentially the core ($\overline{f}\rightarrow
\eta_{0}$) and the tail ($\overline{f}\rightarrow 0$) of the
distribution. Note that the confinement of the tail
($\overline{f}\rightarrow 0$) is consistent with the notion of maximum
entropy ``bubble'' \cite{JFM2}. However, mixing can also be incomplete
inside the ``bubble'', in particular in the core ($\overline{f}\rightarrow
\eta_{0}$) of the distribution.

(iii) {\it Incomplete relaxation in time:} during violent relaxation,
the system tends to approach the statistical equilibrium state
(\ref{E15}). However, as it approaches equilibrium, the fluctuations
of the potential, which are the engine of the evolution, become less
and less effective to drive the relaxation. This is because the scale
of the fluctuations becomes smaller and smaller as time goes on. This
effect can be taken into account in the kinetic theory by considering
that the correlation lengths $\epsilon_{r}(t)$ and $\epsilon_{v}(t)$
decrease with time so that, in the kinetic equation (\ref{v10}), the
prefactor $\epsilon_{r}(t)\epsilon_{v}(t)\rightarrow 0$ for
$t\rightarrow +\infty$.  As a result, the ``turbulent'' current ${\bf
J}$ in Eq. (\ref{v10}) can vanish {\it before} the system has reached
the statistical equilibrium state (\ref{E15}). In that case, the
system can be trapped in a QSS that is a steady solution of the Vlasov
equation different from the statistical prediction (\ref{E15}).

Similar arguments have been given in \cite{rr,csr} on the basis of a more
phenomenological kinetic theory of violent relaxation based on the
MEPP. The idea is to keep the Lynden-Bell entropy (\ref{E9}) unchanged
(as being the fundamental entropy for the process of violent
relaxation) but describe the dynamical evolution of $\rho({\bf r},{\bf
v},\eta,t)$ by a relaxation equation of the form
\begin{equation}
{\partial\rho\over\partial t}+{\bf v}\cdot {\partial\rho\over\partial {\bf r}}-\nabla\Phi\cdot {\partial\rho\over\partial {\bf v}}={\partial\over\partial {\bf v}}\cdot \biggl\lbrace D({\bf r},{\bf v},t)\biggl\lbrack {\partial\rho\over\partial {\bf v}}+\beta(t)(\eta-\overline{f})\rho {\bf v}\biggr\rbrack\biggr\rbrace,
\label{relaxcgw}
\end{equation}
with a diffusion coefficient $D({\bf r},{\bf v},t)$ going to zero for
large time (as the variations of the potential $\Phi$ decay) and in
regions of phase-space where the fluctuations $\delta\Phi$ are not
strong enough to provide efficient mixing (so that $f_{2}\rightarrow
0$). The vanishing of the diffusion coefficient can ``freeze'' the
system in a subdomain of phase space and account for incomplete
relaxation and non-ergodicity. In general, the resulting state,
although incompletely mixed, is not a Tsallis $q$-distribution. This
approach in interesting because it is not based on a generalized
entropy, so there is no free parameter like $q$. However,
it demands to solve a dynamical equation (\ref{relaxcgw}) to predict
the QSS. {\it The idea is that, in case of incomplete
relaxation (non-ergodicity), the prediction of the QSS
is impossible without considering the dynamics} \cite{super}.

\section{9. Generalized $H$-functions in $\overline{f}$-space}

In order to quantify the importance of mixing, Tremaine {\it et al.} 
\cite{thlb} have introduced the notion of generalized $H$-functions. A
generalized $H$-function is a functional of the coarse-grained DF of
the form
\begin{equation}
H[\overline{f}]=-\int C(\overline{f}) d{\bf r}d{\bf v}, \label{i1}
\end{equation}
where $C$ is any convex function ($C''>0$).  We assume that the
initial condition at $t=0$ has been prepared without small-scale
structure so that the fine-grained and coarse-grained DF are equal: $\overline{f}({\bf r},{\bf v},0)=f({\bf r},{\bf v},0)$. For
$t>0$, the system will mix in a complicated manner and develop
intermingled filaments so that these two fields will not be equal
anymore. We have
\begin{eqnarray}
H(t)-H(0)=\int \lbrace C\lbrack \overline{f}({\bf r},{\bf v},0)\rbrack - C\lbrack \overline{f}({\bf r},{\bf v},t)\rbrack \rbrace d{\bf r}d{\bf v}\nonumber\\
=\int \lbrace C\lbrack {f}({\bf r},{\bf v},0)\rbrack - C\lbrack \overline{f}({\bf r},{\bf v},t)\rbrack \rbrace d{\bf r}d{\bf v}.
\label{ha2}
\end{eqnarray}
The fine-grained DF is solution of the Vlasov equation
\begin{eqnarray}
{\partial f\over\partial t}+{\bf v}{\partial f\over\partial
{\bf r}}-\nabla\Phi {\partial f\over \partial
{\bf v}}=0.
\label{ha3}
\end{eqnarray}
Thus
\begin{eqnarray}
{d\over dt}\int C(f)d{\bf r}d{\bf v}=\int C'(f){\partial f\over\partial t}d{\bf r}d{\bf v}=-\int C'(f)\left ({\bf v}\frac{\partial f}{\partial {\bf r}}-\nabla\Phi \frac{\partial f}{\partial {\bf v}}\right )d{\bf r}d{\bf v}\nonumber\\
=-\int \left ({\bf v}\frac{\partial}{\partial {\bf r}}-\nabla\Phi \frac{\partial}{\partial {\bf v}}\right )C(f)d{\bf r}d{\bf v}=0.
\label{ha4}
\end{eqnarray}
This shows that the functional $H\lbrack f\rbrack$ calculated
with the fine-grained DF is independent on time (it is a
particular Casimir) so Eq. (\ref{ha2}) becomes
\begin{eqnarray}
H(t)-H(0)=\int \lbrace C\lbrack {f}({\bf r},{\bf v},t)\rbrack - C\lbrack \overline{f}({\bf r},{\bf v},t)\rbrack \rbrace d{\bf r}d{\bf v}.
\label{ha5}
\end{eqnarray}
Let us now divide a macrocell of surface $\Delta$ into $\nu$
microcells. We call $f_{i}$ the value of the DF in a microcell. The
contribution of a macrocell to $H(t)-H(0)$ is
\begin{eqnarray}
\Delta\biggl\lbrace {1\over\nu}\sum_{i}C(f_{i})- C\biggl ({1\over\nu}\sum_{i} f_{i}\biggr )\biggr\rbrace,
\label{ha6}
\end{eqnarray}
which is positive since $C$ is convex. This implies that
$H(t)-H(0)\ge 0$. We conclude that the generalized $H$-functions
$H\lbrack\overline{f}\rbrack$ calculated with the coarse-grained DF
increase in the sense that $H(t)\ge H(0)$ for any $t\ge 0$
\cite{thlb}. This is similar to the Boltzmann $H$-theorem in kinetic
theory.  However, contrary to the Boltzmann equation, the Vlasov
equation does not single out a unique functional (the above inequality
is true for {\it all} $H$-functions) and the time evolution of the
$H$-functions is not necessarily monotonic (nothing is implied
concerning the relative values of $H(t)$ and $H(t')$ for $t,t'>0$).

{\bf Exemple}: the  Tsallis
functional expressed in terms of the coarse-grained DF
\begin{equation}
H_{q}[\overline{f}]=-{1\over q-1}\int (\overline{f}^{q}-\overline{f})
d{\bf r}d{\bf v},
\label{tsal1h}
\end{equation}
is a particular generalized $H$-function. This interpretation of
Tsallis functional as a generalized $H$-function in the sense of
\cite{thlb} was given by Plastino \& Plastino \cite{pp} and
 Chavanis \cite{pre,super,aa3}. The Boltzmann functional $H_{1}[\overline{f}]=-\int \overline{f}\ln\overline{f} d{\bf r}d{\bf v}$, corresponding to $q=1$ is also a particular $H$-function in the sense of \cite{thlb}.

\section{10. Formal nonlinear dynamical stability for the Vlasov equation}

In this section, we discuss the formal nonlinear dynamical stability of
stationary solutions of the Vlasov equation. We consider a special
class of Casimir functionals of the form
\begin{equation}
S[{f}]=-\int C({f}) d{\bf r}d{\bf v},  \label{i1cas}
\end{equation}
where $C$ is any convex function ($C''>0$). These functionals, as well
as the energy and the mass, are conserved by the Vlasov equation. It
results that the maximization problem
\begin{eqnarray}
\label{dynm}
\max_{f} \quad \lbrace S[{f}]\quad |\quad E[{f}]=E, \ M[{f}]=M \rbrace
\end{eqnarray}
determines a steady state of the Vlasov equation that is formally
nonlinearly dynamically stable. Writing the first variations as
$\delta S-\beta\delta E-\alpha\delta M=0$, the steady state satisfies
$f=F(\beta\epsilon+\alpha)$ where $F(x)=(C')^{-1}(-x)$. Therefore,
this distribution function is a monotonically decreasing function of
the energy (for $\beta>0$): $f=f(\epsilon)$ with
$f'(\epsilon)<0$. Formally, the maximization problem (\ref{dynm}) is
similar to a condition of ``microcanonical stability'' in
thermodynamics where $S$ plays the role of an entropy. Due to this
{\it thermodynamical analogy} \cite{pre}, we can use the methods of
thermodynamics to study the nonlinear dynamical stability problem
(\ref{dynm}). We also note that if ${f}({\bf r},{\bf v})$ solves the
minimization problem
\begin{equation}
\label{na2}
\min_{f} \left\lbrace F[f]=E[f]-TS[f] \ | M[f]=M \right\rbrace,
\end{equation}
then it solves the maximization problem (\ref{dynm}) and is therefore
nonlinearly dynamically stable. However, the reciprocal is wrong in
case of ensemble inequivalence. A solution of (\ref{dynm}) is not
necessarily a solution of (\ref{na2}). Therefore, the minimization
problem (\ref{na2}) provides just a {\it sufficient} condition of
nonlinear dynamical stability. The stability criterion (\ref{dynm}) is
more refined than the stability criterion (\ref{na2}): if a
distribution $f=f(\epsilon)$ with $f'(\epsilon)<0$ satisfies
(\ref{dynm}) or (\ref{na2}), then it is nonlinearly dynamically
stable; however, if is does not satisfy (\ref{na2}) it can be
nonlinearly dynamically stable provided that it satisfies
(\ref{dynm}). This means that we can ``miss'' stable solutions if we
just use the optimization problem (\ref{na2}) to construct stable
stationary solutions of the Vlasov equation of the form
$f=f(\epsilon)$ with $f'(\epsilon)<0$.  The criterion (\ref{na2})
determines a {\it subclass} of solutions of the maximization problem
(\ref{dynm}). The criterion (\ref{dynm}) is richer and allows to
construct a larger class of nonlinearly dynamically stable stationary
solutions.  This is similar to a situation of ``ensemble
inequivalence'' in thermodynamics \cite{ellis,bb}. Indeed, (\ref{na2})
is similar to a criterion of ``canonical stability'' in a
thermodynamical analogy \cite{pre}, where $F$ is similar to a ``free
energy''. Canonical stability implies microcanonical stability but the
converse is wrong in case of ensemble inequivalence.

For a long time, most studies related to the formal nonlinear
dynamical stability of the Vlasov equation considered the minimization
of an Energy-Casimir functional $E-S$ at fixed mass (problem with one
constraint) \cite{holm}. This Energy-Casimir method was first
introduced by Arnold \cite{arnold} in two-dimenional hydrodynamics for
the Euler equation. This corresponds to the ``canonical'' criterion
(\ref{na2}).  It only provides a {\it sufficient} condition of formal
nonlinear dynamical stability. Recently, Ellis {\it et al.}
\cite{ellisgeo} introduced a {\it refined stability criterion} for 
the 2D Euler equation in the form the ``microcanonical'' criterion
(\ref{dynm}) (problem with two constraints). In \cite{aaantonov}, we
introduced a similar criterion to study the formal nonlinear dynamical
stability of stellar systems with respect to the Vlasov-Poisson system.
 Explicit examples of ``ensemble inequivalence'' in the context of
the nonlinear dynamical stability for the 2D Euler and Vlasov equations, where
the criteria (\ref{dynm}) and (\ref{na2}) do not coincide, were
explicitly constructed by Ellis {\it et al.} \cite{ellisgeo} for
geophysical flows and by Chavanis \cite{aaantonov} for stellar
polytropes. In the astrophysical context, it was shown
\cite{aaantonov} that ``ensemble inequivalence'' leads to a nonlinear
version of the Antonov first law (see Sec. 11).

{\bf Example:} the extremization of the Tsallis functional
\begin{equation}
S_{q}[{f}]=-{1\over q-1}\int ({f}^{q}-{f}) d{\bf r}d{\bf v},
\label{tsal1}
\end{equation}
at fixed mass and energy leads to distribution functions of the form
\begin{equation}
\label{q} {f}({\bf r},{\bf v})=\biggl\lbrack \mu-{\beta(q-1)\over
q}\epsilon\biggr\rbrack^{1\over q-1},
\end{equation}
where $\mu$ and $\beta$ are Lagrange multipliers determined by $M$ and
$E$ \cite{lang}. These DF are particular stationary solutions of the
Vlasov equation called stellar polytropes in astrophysics
\cite{bt}. Furthermore, if the DF {\it maximizes} the Tsallis
functional at fixed mass and energy, then it is formally nonlinearly
dynamically stable with respect to the Vlasov equation according to
(\ref{dynm}). In this context, the Tsallis functional (\ref{tsal1}) is
a particular Casimir functional of the form (\ref{i1cas}), not a
generalized entropy. Its maximization at fixed mass and energy forms a
criterion of nonlinear dynamical stability for the Vlasov equation,
not a criterion of generalized thermodynamical stability in the
microcanonical ensemble. This dynamical interpretation of the Tsallis
functional related to the Vlasov equation was given by Chavanis \&
Sire \cite{cstsallis}. It differs from the thermodynamical
interpretation given by Taruya \& Sakagami
\cite{ts} in terms of generalized thermodynamics. On the other hand,
the minimization of the Tsallis functional $F_{q}=E-TS_{q}$ at fixed
mass forms a {\it sufficient} criterion of nonlinear dynamical
stability for the Vlasov equation, not a criterion of generalized
thermodynamical stability in the canonical ensemble. In the present
context, the resemblence with a generalized thermodynamical formalism
is effective and is the mark of a {\it thermodynamical analogy}
\cite{pre}. Note that the preceding discussion also applies to the
Boltzmann distribution, corresponding to $q=1$, which is also a
particular solution of the Vlasov equation.

{\bf Numerical algorithms:} A relaxation equation has been proposed to
solve the maximization problem (\ref{dynm}). It is obtained from a
Maximum Entropy Production Principle (MEPP), exploiting the
thermodynamical analogy, and has the form of a generalized mean field
Kramers equation \cite{pre}. It can serve as a numerical algorithm
to determine a nonlinearly dynamically stable stationary solution of
the Vlasov equation, specified by the convex function $C(f)$, for
given $E$ and $M$. The relaxation equation solving the
``microcanonical'' problem (\ref{dynm}) is
\begin{equation}
{\partial {f}\over\partial t}+{\bf v}\cdot {\partial {f}\over\partial {\bf r}}-\nabla\Phi\cdot {\partial {f}\over\partial {\bf v}}={\partial\over\partial {\bf v}}\cdot \biggl\lbrace D\biggl\lbrack {\partial {f}\over\partial {\bf v}}+\frac{\beta(t)}{C''({f})} {\bf v}\biggr\rbrack\biggr\rbrace,
\label{p13}
\end{equation}
with
\begin{equation}
\beta(t)=-\frac{\int D\frac{\partial {f}}{\partial {\bf v}}\cdot {\bf v} d{\bf r}d{\bf v}}{\int D \frac{v^{2}}{C''(f)}d{\bf r}d{\bf v}}.
\label{p13b}
\end{equation}
This equation conserves the mass $M$, the energy $E$ (through the time
dependent Lagrange multiplier $\beta(t)$) and monotonically increases
the $S$-functional: $\dot S\ge 0$ ($H$-theorem). Furthermore, $\dot
S=0$ iff $f$ is a steady solution of Eq. (\ref{p13}). Therefore,
$S[f]$ is the Lyapunov functional of the relaxation equation
(\ref{p13})-(\ref{p13b}). It results that $f({\bf r},{\bf v})$ is a
linearly dynamically stable stationary solution of
Eq. (\ref{p13})-(\ref{p13b}) iff it is a (local) maximum of $S$ at
fixed $E$, $M$. Minima or saddle points of $S$ are dynamically
unstable. Furthermore, if $S$ is bounded from above (within the
previous constraints), $f({\bf r},{\bf v},t)$ will relax towards a
(local) maximum of $S$ at fixed $E$, $M$ for $t\rightarrow +\infty$
(if there exists several local maxima of $S$, the selection will
depend on a complicated notion of basin of attraction). Therefore,
Eq. (\ref{p13})-(\ref{p13b}) can serve as a numerical algorithm to
solve the ``microcanonical'' problem (\ref{dynm}) and compute a
nonlinearly dynamically stable stationary solution of the Vlasov
equation, specified by $C$, for given $E$ and $M$.  Note that the
generalized Vlasov-Landau (GVL) equation
\begin{equation}
\label{bol13q} {\partial f\over\partial t}+{\bf v}\cdot
{\partial f\over\partial {\bf r}}-\nabla\Phi \cdot {\partial
f\over\partial {\bf v}}=A{\partial\over\partial
v^{\mu}}\int \frac{\delta^{\mu\nu}w^2-w^{\mu}w^{\nu}}{w^{3}} \biggl (\frac{1}{C''(f_{1})}{\partial
f\over\partial v^{\nu}}-\frac{1}{C''(f)}{\partial
f_{1}\over\partial v_{1}^{\nu}}\biggr )d{\bf v}_{1},
\end{equation}
introduced in
\cite{genlandau} has the same properties as the relaxation equation (\ref{p13})-(\ref{p13b}). In particular,  {\it $f({\bf r},{\bf v})$ is a
linearly dynamically stable stationary solution of the GVL equation
iff it is a (local) maximum of $S$ at fixed $E$, $M$}. Minima or
saddle points of $S$ are dynamically unstable. Therefore, the GVL
equation can also provide a numerical algorithm to compute nonlinearly
dynamically stable stationary solutions of the Vlasov equation,
specified by $C$, for given $E$ and $M$. We note that {\it $f({\bf
r},{\bf v})$ is a linearly dynamically stable stationary solution of
the generalized Vlasov-Landau equation iff it is a formally
nonlinearly dynamically stable stationary solution of the Vlasov
equation.}

If we fix the inverse temperature $\beta$ instead of the energy in
Eq. (\ref{p13}), we get a relaxation equation
\begin{equation}
{\partial {f}\over\partial t}+{\bf v}\cdot {\partial {f}\over\partial {\bf r}}-\nabla\Phi\cdot {\partial {f}\over\partial {\bf v}}={\partial\over\partial {\bf v}}\cdot \biggl\lbrace D\biggl\lbrack {\partial {f}\over\partial {\bf v}}+\frac{\beta}{C''({f})} {\bf v}\biggr\rbrack\biggr\rbrace,
\label{p13c}
\end{equation}
solving the ``canonical'' problem (\ref{na2}). It is called the
generalized Vlasov-Kramers (GVK) equation. This equation conserves the
mass $M$ and monotonically decreases the $F$-functional: $\dot F\le
0$. Furthermore, $\dot F=0$ iff $f$ is a steady solution of
Eq. (\ref{p13c}). Therefore, $F[f]$ is the Lyapunov functional of the
GVK equation. It results that {\it $f({\bf r},{\bf v})$ is a linearly
dynamically stable stationary solution of the GVK equation iff it is a
(local) minimum of $F$ at fixed $M$}. Maxima or saddle points of $F$
are dynamically unstable. Furthermore, if $F$ is bounded from below
(within the previous constraints), $f({\bf r},{\bf v},t)$ will relax
towards a (local) minimum of $F$ at fixed $M$. Therefore,
Eq. (\ref{p13c}) can serve as a numerical algorithm to solve the
``canonical problem'' (\ref{na2}) and determine a subclass of
nonlinearly dynamically stable stationary solutions of the Vlasov
equation, specified by $C$, since (\ref{na2}) $\Rightarrow$
(\ref{dynm}). We note that {\it a linearly dynamically stable
stationary solution of the generalized Vlasov-Kramers equation is (i)
a formally nonlinearly dynamically stable stationary solution of the
Vlasov equation (ii) a linearly dynamically stable stationary solution
of the generalized Vlasov-Landau equation (but the reciprocal is wrong in case of ensemble inequivalence).}

We can also use the preceding relaxation equations to solve the
optimization problems (\ref{projm}) and (\ref{projc}) of Sec. 5. If we
view $f$ as the coarse-grained DF $\overline{f}$ and if we take
$C(\overline{f})$ of the form (\ref{rel}), a linearly dynamically
stable stationary solution of the relaxation equation
(\ref{p13})-(\ref{p13b}) is a Lynden-Bell statistical equilibrium
state according to the sufficient condition (\ref{projm})
$\Rightarrow$ (\ref{mlbm}).  Similarly, {\it a linearly dynamically
stable stationary solution of the generalized Vlasov-Landau equation
is a Lynden-Bell statistical equilibrium state (but the reciprocal is
wrong in case of ensemble inequivalence).} On the other hand, a
linearly dynamically stable stationary solution of the relaxation
equation (\ref{p13c}) is a Lynden-Bell statistical equilibrium state
according to the sufficient condition (\ref{projc}) $\Rightarrow$
(\ref{mlbm}). Therefore, {\it a linearly dynamically stable stationary
solution of the generalized Vlasov-Kramers equation is a Lynden-Bell
statistical equilibrium state (but the reciprocal is wrong in case of
ensemble inequivalence).}

\section{11. The nonlinear Antonov first law}

It was first realized by Antonov \cite{antolaw} in astrophysics, that
there exists deep connections between the dynamical stability of
steady solutions of the Vlasov equation (\ref{vp1}) of the form
$f=f(\epsilon)$ with $f'(\epsilon)<0$ and the dynamical stability of
steady solutions of the Euler equations
\begin{equation}
\label{vp1gr}
{\partial \rho\over\partial t}+\nabla\cdot (\rho {\bf u})=0,
\end{equation}
\begin{equation}
\label{vp1kh}
\frac{\partial {\bf u}}{\partial t}+({\bf u}\cdot \nabla){\bf u}=-\frac{1}{\rho}\nabla p-\nabla\Phi,
\end{equation}
with a barotropic equation of state $p=p(\rho)$ (as usual $\Phi({\bf
r},t)=\int u(|{\bf r}-{\bf r}'|)\rho({\bf r}',t)d{\bf r}'$ is the mean
field potential). The Euler equations conserve the mass $M[\rho]=\int
\rho d{\bf r}$ and the energy
\begin{eqnarray}
{\cal W}[\rho,{\bf u}]=\int \rho \int^{\rho}{p(\rho')\over\rho'^{2}}d\rho'
d{\bf r}+{1\over
  2}\int\rho\Phi d{\bf r}+\frac{1}{2}\int\rho {\bf u}^{2}d{\bf r},
\label{fef15}
\end{eqnarray}
including the internal energy, the potential energy and the kinetic
energy of the mean motion.  It results that the minimization problem
\begin{eqnarray}
\label{dyneuler}
\min_{\rho,{\bf u}}\quad \lbrace {\cal W}[{\rho},{\bf u}]\quad |\quad  M[{\rho}]=M \rbrace,
\end{eqnarray}
determines a steady state of the Euler equation that is formally nonlinearly
dynamically stable. Writing the first order variations as $\delta {\cal W}-\alpha\delta M=0$, we get ${\bf u}={\bf 0}$ and
\begin{eqnarray}
\label{dlr}
\int^{\rho} \frac{p'(\rho')}{\rho'}d\rho'=-\Phi.
\end{eqnarray}
This leads to the condition of hydrostatic equilibrium
\begin{eqnarray}
\label{few}
\nabla p+\rho\nabla\Phi={\bf 0}.
\end{eqnarray}
We also note that $\rho=\rho(\Phi)$ and that
$p'(\rho)/\rho=-1/\rho'(\Phi)$ so that $\rho(\Phi)$ is monotonically
decreasing (since $p'(\rho)>0$ in cases of physical interest). 

Let us now return to the Vlasov equation.  We have seen that the
minimization problem (\ref{na2}) provides a {\it sufficient} condition
of formal nonlinear dynamical stability. In order to solve this minimization
problem, we can proceed in two steps (we consider here the global
minimization problem. The local minimization problem is treated in the
Appendix). {\it First step:} we determine the distribution $f_1({\bf
r},{\bf v})$ which minimizes $F[f]$ at fixed density profile
$\rho({\bf r})=\int fd{\bf v}$.  This gives a distribution
$f_1[\rho({\bf r}),{\bf v}]$ depending on $\rho({\bf r})$ and ${\bf
v}$. Substituting this distribution in the functional $F[f]$, we
obtain a functional $F[\rho]\equiv F[f_1]$ of the density. {\it Second
step:} we determine the density $\rho_*({\bf r})$ which minimizes
$F[\rho]$ at fixed mass $M[\rho]=M$. Finally, we have $f_*({\bf
r},{\bf v})=f_1[\rho_*({\bf r}),{\bf v}]$.  Let us be more
explicit. If we fix the density profile $\rho({\bf r})$, the potential
energy $W[\rho]$ is automatically determined. Therefore, minimizing
$F[f]=E[f]-TS[f]$ at fixed density profile is equivalent to minimizing
$\tilde{F}[f]=K[f]-TS[f]$ at fixed density profile, where
$K[f]=\frac{1}{2}\int f v^{2}d{\bf r}d{\bf v}$ is the kinetic
energy. The distribution $f_1({\bf r},{\bf v})$ that extremizes
$\tilde{F}[f]$ with the constraint $\int f \, d{\bf v} =\rho({\bf r})$
satisfies the first order variations $\delta F+\int
\lambda({\bf r}) \delta \int f d{\bf v} d{\bf r}=0$, where
$\lambda({\bf r})$ is a Lagrange multiplier. This leads to
\begin{equation}
\label{fef3} f_{1}=F\left\lbrace \beta\biggl\lbrack {v^{2}\over
2}+\lambda({\bf r})\biggr\rbrack\right\rbrace,
\end{equation}
where $F(x)=(C')^{-1}(-x)$ and $\lambda({\bf r})$ is determined by $\rho({\bf
r})$ by writing $\rho=\int f_{1}d{\bf v}$. Since
\begin{equation}
\label{fef3b}\delta^{2}F=-T\delta^{2}S=\frac{1}{2}T\int C''(f_{1})(\delta f)^{2}d{\bf r}d{\bf v}\ge 0,
\end{equation}
this is a {\it minimum} of $F[f]$ at fixed density profile. According
to Eq. (\ref{fef3}), the density $\rho=\int f_{1}d{\bf
v}=\rho[\lambda({\bf r})]$ and the pressure $p=\frac{1}{d}\int
f_{1}v^{2}d{\bf v}=p[\lambda({\bf r})]$ are functions of $\lambda({\bf
r})$. Eliminating $\lambda({\bf r})$ between these two expressions we
obtain a barotropic equation of state $p=p(\rho)$. Then, after simple
calculations (see Appendix B of \cite{aaantonov}), we can show that
the functional $F[\rho]\equiv F[f_1]$ is given by
\begin{eqnarray}
{F}[\rho]=\int \rho \int^{\rho}{p(\rho')\over\rho'^{2}}d\rho'
d{\bf r}+{1\over
  2}\int\rho\Phi d{\bf r}.
\label{fef15b}
\end{eqnarray}
Therefore, we conclude that $f_{*}({\bf r},{\bf v})=f_1[\rho_*({\bf
r}),{\bf v}]$ solves the minimization problem (\ref{na2}) iff
$\rho_{*}({\bf r})$ solves the minimization problem
\begin{eqnarray}
\label{dyneq}
\min_{\rho} \quad \lbrace F[{\rho}]\quad |\quad M[{\rho}]=M\rbrace.
\end{eqnarray}
Now we observe that the solutions of the minimization problems
(\ref{dyneuler}) and (\ref{dyneq}) coincide. Therefore, we conclude
that $f_{*}({\bf r},{\bf v})=f_1[\rho_*({\bf r}),{\bf v}]$ solves the
minimization problem (\ref{na2}) iff $\rho_{*}({\bf r})$ solves the
minimization problem (\ref{dyneuler}). On the other hand, since the
ensemble of solutions of (\ref{na2}) is included in the ensemble of
solutions of (\ref{dynm}) we conclude that : $f_{*}({\bf r},{\bf
v})=f_1[\rho_*({\bf r}),{\bf v}]$ solves the maximization problem
(\ref{dynm}) if $\rho_{*}({\bf r})$ solves the minimization problem
(\ref{dyneuler}), but the reciprocal is wrong in case of ensemble
inequivalence.  In conclusion, we find that a DF $f=f(\epsilon)$ with
$f'(\epsilon)<0$ is a nonlinearly dynamically stable steady state of
the Vlasov equation (kinetic system) if the corresponding density
profile $\rho({\bf r})$ is a nonlinearly dynamically stable steady
state of the Euler equation (fluid system), but the reciprocal is
wrong. In astrophysics, this forms the nonlinear Antonov first law
\cite{aaantonov}: ``a stellar system with $f=f(\epsilon)$ and
$f'(\epsilon)<0$ is a nonlinearly dynamically stable steady state of
the Vlasov-Poisson system if the corresponding barotropic star is a
nonlinearly dynamically stable steady state of the Euler-Poisson
system (but the reciprocal is wrong).''
\footnote{This law was proven by Antonov \cite{antolaw} 
for the problem of {\it linear} dynamical stability and extended by
Chavanis
\cite{aaantonov} to the problem of formal nonlinear dynamical
stability.} This can be seen as a manifestation of ``ensemble
inequivalence'' in a thermodynamical analogy
\cite{aaantonov}. For example, complete polytropes with index
$3<n<5$ satisfy the ``microcanonical'' criterion (\ref{dynm}) but not
the ``canonical'' criterion (\ref{na2}) $\Leftrightarrow$ (\ref{dyneq})
$\Leftrightarrow$ (\ref{dyneuler}). Thus, stellar polytropes with index
$3<n<5$ are nonlinearly dynamically stable with respect to the
Vlasov-Poisson system while the corresponding gaseous polytropes are
{\it unstable} with respect to the Euler-Poisson system \cite{aaantonov}.

{\bf Example:} When $S[f]$ is the Tsallis functional (\ref{tsal1}),
the corresponding barotropic gas is a gaseous polytrope with an
equation of state \cite{cstsallis}:
\begin{eqnarray}
p=K\rho^{\gamma}, \qquad \gamma=1+\frac{1}{n}, \qquad n=\frac{1}{q-1}+\frac{d}{2}.
\label{fef15c}
\end{eqnarray}
The functional (\ref{fef15b}) is of the form
\begin{eqnarray}
{F}[\rho]=\frac{K}{\gamma-1}\int (\rho^{\gamma}-\rho) d{\bf r}+{1\over
  2}\int\rho\Phi d{\bf r}.
\label{fef15d}
\end{eqnarray}
This can be viewed as a Tsallis ``free energy'' functional
$F_{\gamma}[\rho]=W[\rho]-KS_{\gamma}[\rho]$ with index $\gamma$ where
$W$ is the (potential) energy and $K$ plays the role of a ``polytropic
temperature'' (recall again that these analogies with thermodynamics
are purely formal in the present {\it dynamical} context). It can be
compared with the Tsallis ``free energy'' functional
$F_{q}[f]=E[f]-TS_{q}[f]$ in phase space (see Sec. 10). The steady
density distribution (\ref{dlr}) is of the form
\begin{eqnarray}
\rho=\left\lbrack \lambda-\frac{\gamma-1}{K\gamma}\Phi\right\rbrack^{\frac{1}{\gamma-1}}
\label{fef15e}
\end{eqnarray}
which is morphologically similar to Eq. (\ref{q}). Therefore, {\it  a Tsallis
distribution $f_{q}(\epsilon)$ in phase space yields a Tsallis
distribution $\rho_{\gamma}(\Phi)$ in position space} \cite{cstsallis}.

{\bf Numerical algorithm:} A relaxation equation has been proposed to
solve the minimization problem (\ref{dyneq}).  It  is called the generalized
Smoluchowski (GS) equation \cite{pre}. It can serve as a numerical algorithm to
determine nonlinearly dynamically stable stationary solutions of the
Euler equations, specified by $p(\rho)$, for
given $M$. It is written as
\begin{equation}
\label{gs}
{\partial \rho\over\partial t}=\nabla\cdot\left\lbrack\frac{1}{\xi}(\nabla p+\rho\nabla\Phi)\right\rbrack.
\end{equation}
This equation conserves the mass $M$ and monotonically increases the
$F$-functional: $\dot F\ge 0$ ($H$-theorem). Furthermore, $\dot F=0$
iff $\rho$ is a steady solution of (\ref{gs}). Therefore, $F[\rho]$ is
the Lyapunov functional of the GS equation (\ref{gs}).  It results
that {\it $\rho({\bf r})$ is a linearly dynamically stable stationary
solution of the generalized Smoluchowski equation iff it is a (local)
minimum of $F$ at fixed $M$.} Maxima or saddle points of $F$ are
dynamically unstable.  Furthermore, if $F$ is bounded from below
(within the previous constraint), $\rho({\bf r},t)$ will relax towards
a (local) minimum of $F$ at fixed $M$ for $t\rightarrow
+\infty$. Therefore, Eq. (\ref{gs}) can serve as a numerical algorithm
to solve the ``canonical'' problem (\ref{dyneq}) and compute a
nonlinearly dynamically stable stationary solution of the Euler
equation, specified by $p(\rho)$, for given $M$ since (\ref{dyneq})
$\Leftrightarrow$ (\ref{dyneuler}). We note that {\it $\rho({\bf r})$
is a linearly dynamically stable stationary solution of the
generalized Smoluchowski equation iff it is a formally nonlinearly
dynamically stable stationary solution of the barotropic Euler
equation}.  There are other corrolaries to that result. Since
(\ref{na2}) $\Leftrightarrow$ (\ref{dyneq}) we conclude that {\it
$f({\bf r},{\bf v})$ is linearly dynamically stable with respect to
the generalized Vlasov-Kramers equation (\ref{p13c}) iff the
corresponding density $\rho({\bf r})$ is linearly dynamically stable
with respect to the generalized Smoluchowski equation (\ref{gs}).} On
the other hand, according to the implications (\ref{dyneq})
$\Leftrightarrow$ (\ref{dyneuler}) $\Leftrightarrow$ (\ref{na2})
$\Rightarrow$ (\ref{dynm}) and (\ref{dyneq}) $\Leftrightarrow$
(\ref{na2}) $\Leftrightarrow$ (\ref{mf1c}) $\Rightarrow$ (\ref{mf1})
$\Rightarrow$ (\ref{mlbm}), we conclude that {\it a linearly
dynamically stable stationary solution of the generalized Smoluchowski
equation determines (i) a nonlinearly dynamically stable stationary
solution of the Vlasov equation (ii) a linearly dynamically stable
stationary solution of the generalized Vlasov-Landau equation
(\ref{bol13q}) (iii) a Lynden-Bell statistical equilibrium state (but
the reciprocal is wrong in case of ensemble inequivalence).}

\section{12. Selective decay principle}

Let us conclude this paper by some phenomenological
considerations. The results of Sec. 9 suggest a notion of generalized
{\it selective decay principle} \cite{super,next05}: among all
invariants of the collisionless dynamics, the generalized
$H$-functions (fragile constraints) increase ($-H$ decrease) on the
coarse-grained scale while the mass and the energy (robust
constraints) are approximately conserved\footnote{A selective decay
principle was first introduced in two-dimensional turbulence, for the
2D Navier-Stokes equation, based on different arguments
\cite{montgo}. It states that, in the presence of a small viscosity
$\nu\rightarrow 0$, the energy of the flow is approximately conserved
while the enstrophy decays. This leads to a phenomenological minimum
enstrophy principle where the ``equilibrium'' flow minimizes the
enstrophy at fixed circulation and energy.  Our selective decay
principle, which also applies to the (inviscid) 2D Euler equation
\cite{super}, is more general (the enstrophy is a {\it particular}
$H$-function) and uses the notion of coarse-graining rather than
viscosity or other source of dissipation.}.  According to this
principle, we {\it may} expect (phenomenologically) that the QSS
reached by the system as a result of violent relaxation will maximize
a certain $H$-function called $H^{*}[\overline{f}]$ (non-universal) at
fixed mass $M$ and energy $E$. Therefore, it is expected to satisfy a
problem of the form
\begin{eqnarray}
\label{seld}
\max_{\overline{f}}\quad \lbrace H^{*}[\overline{f}]\quad |\quad E[\overline{f}]=E, M[\overline{f}]=M\rbrace. \qquad ({\rm Phenomenological})
\end{eqnarray}
As we have seen, this maximization problem determines a distribution
function $\overline{f}=\overline{f}(\epsilon)$ with
$\overline{f}'(\epsilon)<0$ which is a stationary solution of the
Vlasov equation (recall that our argument applies to the {\it
coarse-grained} distribution). In general, the $H$-function
$H^{*}[\overline{f}]$ that is effectively maximized by the system as a
result of violent relaxation (if it really maximizes an $H$-function!) 
is difficult to predict.  It depends on the initial conditions {\it
and} on the efficiency of mixing. If mixing is complete (ergodicity),
the statistical theory of Lynden-Bell shows that the coarse-grained DF
(\ref{E13}) extremizes a functional of the form (\ref{g2})-(\ref{g10})
at fixed mass and energy. Although we cannot conclude that this
extremum is always a maximum because of the problem of ensemble
inequivalence reported in Sec. 5, it is likely however that in many
cases the coarse-grained DF predicted by the statistical theory of
Lynden-Bell maximizes the generalized entropy (\ref{fo})-(\ref{rel})
at fixed mass and energy. Therefore, we conclude that when mixing is
complete $H^{*}=S$, where $S$ is given by
Eqs. (\ref{fo})-(\ref{rel}). In other words, apart from the problem of
ensemble inequivalence which demands further investigation, the
statistical theory of Lynden-Bell {\it justifies} the generalized
selective decay principle (\ref{seld}) and gives the expression of
$H^{*}[\overline{f}]$, depending on the initial condition. We again
emphasize the importance of the notion of coarse-graining: the
functionals $H[f]$ calculated with the fine-grained DF are conserved
while the $H$-functions $H[\overline{f}]$ calculated with the
coarse-grained DF increase. Furthermore, the particular $H$-function
$H^{*}[\overline{f}]=S[\overline{f}]$ (whose form depends on the
initial conditions) is maximum at statistical equilibrium while
$E[\overline{f}]$ and $M[\overline{f}]$ are approximately
conserved. {\it Therefore, although the Lynden-Bell theory takes into
account the conservation of the Casimirs $H[f]$, it leads to a
coarse-grained distribution which maximizes a functional
$H^{*}[\overline{f}]$ at fixed $E$ and $M$}. On the other hand, if
mixing is incomplete (see Sec. 6), the coarse-grained DF may still
maximize a certain generalized $H$-function at fixed mass and energy
but $H^{*}[\overline{f}]$ and $\overline{f}(\epsilon)$ can take forms
that are {\it not} compatible with the expressions
(\ref{fo})-(\ref{rel}) and (\ref{E14}) derived in the statistical
approach. In case of incomplete relaxation, the prediction of this
functional $H^{*}[\overline{f}]$ seems out-of-reach because its form
depends on the efficiency of mixing which is not known {\it a priori}.
In particular, there is no reason why $H^{*}[\overline{f}]$ should
always (universaly) be the Tsallis functional. However, it has been
observed in several occasions \cite{boghosian,campa,paper3} that the
Tsallis distributions are ``attractors'' of the solutions of the
(coarse-grained) Vlasov equation for some initial
conditions.

We now observe that the criterion of formal nonlinear dynamical
stability (\ref{dynm}) is remarkably consistent with the
phenomenological selective decay principle of violent relaxation
(\ref{seld}) provided that we interprete $f$ as the {\it
coarse-grained} DF. This coincidence looks surprising at first
sight. A priori, the optimization problems (\ref{dynm}) and
(\ref{seld}) are completely different: the problem (\ref{dynm})
explicitly uses the fact that the functional $H[f]$ is conserved by
the Vlasov equation while the problem (\ref{seld}) uses the fact that
the functional $H[\overline{f}]$ calculated with the coarse-grained DF
increases during violent relaxation. In fact, the problem (\ref{dynm})
indicates that {\it if} a DF is a maximum of $H$ at fixed mass and
energy, then it is nonlinearly dynamically stable with respect to the
Vlasov equation while the phenomenology of violent relaxation leading
to the problem (\ref{seld}) explains {\it how}, starting from an
unstable initial condition, a DF can possibly reach a maximum of $H$
at fixed mass and energy (through mixing) although $H$ is rigorously
conserved by the Vlasov equation. Indeed, during mixing the
coarse-grained DF is not conserved by the Vlasov equation
($D\overline{f}/Dt\neq 0$ where $D/Dt$ is the material derivative in
phase space) and the $H$-functions $H[\overline{f}]$ increase. Once it
has mixed, the coarse-grained DF satisfies the Vlasov equation
$D\overline{f}/Dt= 0$ so that the Casimirs calculated with the
coarse-grained DF are now conserved $\dot H[\overline{f}]=0$, as well
as the mass and the energy. Indeed, after mixing the coarse-grained
field must be viewed as a new fine-grained field in a possibly further
evolution.  Now, the phenomenology of violent relaxation (\ref{seld})
suggests that $\overline{f}({\bf r},{\bf v},t)$ will be brought to a
maximum $\overline{f}_{*}({\bf r},{\bf v})$ of a certain $H$-function
$H^{*}$. This is {\it proven} by the
theory of Lynden-Bell in case of efficient mixing (up to the problem
of ensemble inequivalence) and this may remain true even in case of
incomplete relaxation. Since $H[\overline{f}]$, $E[\overline{f}]$ and
$M[\overline{f}]$ are conserved by the Vlasov equation after mixing,
and since $\overline{f}_{*}({\bf r},{\bf v})$ is a maximum of
$H^{*}[\overline{f}]$ at fixed mass and energy, then it is a formally
nonlinearly dynamically stable steady state of the Vlasov equation
according to the criterion (\ref{dynm}).

Let us make a final comment. The maximization (under appropriate
constraints) of the Lynden-Bell entropy (\ref{E9}), of the Tsallis
entropy (\ref{se6b}), of an H-function (\ref{i1}) or of a convex
Casimir (\ref{i1cas}) leads to a distribution function of the form
$\overline{f}=\overline{f}(\epsilon)$ with $\overline{f}'(\epsilon)<0$
depending only on the energy. In astrophysics, these DF can only
describe {\it spherical} stellar systems (and even a sub-class of
them) \cite{bt}. In reality, stellar systems are not spherical and
their distribution functions are not function of the energy alone
\footnote{This shows that the phenomenological principle (\ref{seld})
is not always valid. Indeed, real (non-spherical) galaxies do {\it
not} maximize an $H$ function at fixed mass and energy although they
probably result from a process of violent relaxation.}. Indeed,
according to the Jeans theorem
\cite{bt}, there exists more general stationary solutions of the
Vlasov equation which depend on other integrals of motion. This
indicates that the structure of the final state of a collisionless
stellar system depends on its dynamical evolution in a complicated
manner. An important problem in astrophysics is therefore to find the
form of distribution function appropriate to real galaxies. Simple
concepts based on entropies, $H$-functions or formal nonlinear
dynamical stability are not sufficient to understand the structure of
real galaxies. This is particularly deceptive. However, conceptually,
the theory of violent relaxation is important to explain {\it how} a
collisionless stellar system reaches a steady state (QSS). This is due
to phase mixing in phase space. The coarse-grained DF
$\overline{f}({\bf r},{\bf v},t)$ reaches a steady state
$\overline{f}_{QSS}({\bf r},{\bf v})$ in a few dynamical times while the
fine-grained distribution function ${f}({\bf r},{\bf v},t)$ develops
filaments at smaller and smaller scales and is never steady
(presumably). Since this mixing process is very complex, the resulting
structure $\overline{f}_{QSS}({\bf r},{\bf v})$ should be extremely robust
and should be therefore a nonlinearly dynamically stable stationary
solution of the Vlasov equation. {\it Thus, the theory of incomplete
violent relaxation explains how collisionless stellar systems can be
trapped in nonlinearly dynamically stable (robust) stationary solutions of the
Vlasov equation on the coarse-grained scale}.

\section{13. Summary: the different functionals}

We would like to again emphasize the distinction between entropies and
$H$-functions \cite{super}. An entropy is a quantity which is
proportional to the logarithm of the disorder, where the disorder is
equal to the number of microstates consistent with a given
macrostate. This is how the Lynden-Bell entropy (\ref{E9}) has been
defined. Tsallis entropy (\ref{se6b}) could be considered as a
generalization of this definition in the case where the phase-space
has a complex structure so that the evolution is non-ergodic and {\it
the potentially accessible microstates are not equiprobable}.  In each
case, the entropy is a functional of the probability $\rho({\bf
r},{\bf v},\eta)$ and the maximization of these entropies at fixed
mass, energy and Casimirs is a condition of thermodynamical stability.
The ``generalized entropies'' (\ref{fo})-(\ref{rel}) defined in Sec. 5
can be regarded as entropies which are proportional to the logarithm
of the number of microstates consistent both with a given macrostate
and with the constraints imposed by the Vlasov equation
(Casimirs). Their functional form depends on the initial
condition. They are defined on a projection space
($\overline{f}$-space) where a macrostate is defined by the
specification of $\overline{f}({\bf r},{\bf v})$ instead of $\rho({\bf
r},{\bf v},\eta)$. On the other hand, the $H$-functions do not have a
statistical origin. They are just arbitrary functionals of the
coarse-grained distribution $\overline{f}({\bf r},{\bf v},t)$ of the
form (\ref{i1}). They increase during violent relaxation and they are
useful to characterize the degree of mixing of a collisionless system
\cite{thlb}. According to the phenomenological selective decay
principle (\ref{seld}), the QSS reached after violent relaxation is
expected to maximize a certain $H$-function (non-universal) at fixed
mass and energy. According to criterion (\ref{dynm}), this principle
ensures that the QSS is nonlinearly dynamically stable with respect to
the Vlasov equation. This is to be expected since the QSS results from
a turbulent mixing which makes it very robust. In case of complete
violent relaxation, the selective decay principle (\ref{seld}) is
justified by the statistical theory of Lynden-Bell and $H^{*}=S$ where
$S$ is given by Eqs. (\ref{fo})-(\ref{rel}). In case of incomplete
relaxation, the $H$-function $H^{*}$ and the distribution function
$\overline{f}_{QSS}$ of the QSS are different from the Lynden-Bell
prediction.  In some cases of incomplete relaxation
\cite{boghosian,brands,campa,paper3}, the QSS is close to a Tsallis
distribution and $H^{*}$ is close to the Tsallis functional, but this
is not universaly true.

{\bf Examples:} Tsallis functional $S_{q}[\rho]$ expressed in terms of
$\rho({\bf r},{\bf v},\eta)$ is a generalized entropy. Tsallis
functional $S_{q}[\overline{f}]$ expressed in terms of
$\overline{f}({\bf r},{\bf v})$ is (i) a generalized $H$-function (ii)
a particular case of generalized entropy $S_{q}[\rho]$ for two levels
$f\in\lbrace 0,\eta_{0}\rbrace$ in the dilute limit $\overline{f}\ll
\eta_0$. Tsallis functional $S_{q}[{f}]$ expressed in terms of
${f}({\bf r},{\bf v})$ is a particular convex Casimir whose
maximization at fixed mass and energy provides a condition of formal
nonlinear dynamical stability for the Vlasov equation (this leads to
stellar polytropes in astrophysics).

\section{14. Conclusion}

In this paper, we have studied the dynamics and thermodynamics of
systems with weak long-range interactions. In particular, we have
stressed the physical interpretation of the different functionals
appearing in the analysis. In our discussion, we have exclusively
considered the case of {\it isolated} Hamiltonian systems described by the
microcanonical ensemble. In this concluding section, we would like to
briefly discuss the physics of related systems.

It is interesting to compare the microcanonical evolution of
Hamiltonian systems with long-range interactions to the canonical
evolution of Brownian systems with long-range interactions
\cite{paper1}. In the case of Brownian particles in interaction, the
microscopic equations of motion consist in $N$ coupled Langevin
equations including a friction force and a stochastic force in
addition to the long-range coupling.  These additional terms can take
into account short-range interactions with an external medium playing
the role of a thermal bath. These Brownian systems are {\it dissipative} and
they are described by the canonical ensemble. In the thermodynamic
limit $N\rightarrow +\infty$ defined in Sec. 2.2., the mean field
approximation is exact and the evolution of the distribution function
$f({\bf r},{\bf v},t)$ of the Brownian system is governed by the mean
field Kramers equation (decreasing monotonically the Boltzmann free
energy)
\cite{paper2}. This is the canonical counterpart of the Vlasov
equation for Hamiltonian systems.  This shows that, for $N\rightarrow
+\infty$, the Brownian system tends to relax towards the mean field
Boltzmann distribution on a typical timescale $\xi^{-1}$ where $\xi$
is the friction coefficient. In the strong friction limit
$\xi\rightarrow +\infty$, the velocity distribution is close to
Maxwellian and the evolution of the spatial density $\rho({\bf r},t)$
is governed by the mean field Smoluchowski equation \cite{paper2}. So
far, two types of Brownian systems with weak long-range interactions
have been studied: (i) self-gravitating Brownian particles (or
bacterial populations) where the particles interact via a Newtonian
potential \cite{virial2} (ii) the Brownian Mean Field (BMF) model
where the particles interact via a cosine potential \cite{cvb,bo}.

The previously described Hamiltonian and Brownian systems
\cite{paper1} are ``simple'' systems (this does not mean that their
study is trivial!)  because everything is contained in the $N$-body
Hamiltonian or Langevin equations. Recently, some authors have
considered the case of ``complex'' systems where the microscopic
dynamics is not perfectly known or can be biased with respect to the
ordinary one because of some microscopic constraints whose influence
is difficult to formalize. It has been proposed that such systems
could be described by an effective generalized thermodynamics
(E.G.T.). The Tsallis
\cite{tsallis} entropy has become very popular to describe complex
systems, but other forms of generalized entropies can also be considered
\cite{pre,nfp}.  In that context, generalized kinetic equations,
associated with generalized forms of entropic functionals, have been
introduced either in the microcanonical ensemble or in the canonical
ensemble. In the first case, they have the form of generalized
Boltzmann or Landau equations \cite{kaniadakis,gl} and they conserve
the energy and the mass. In the second case, they have the form of
generalized Fokker-Planck equations, like the generalized Kramers and
Smoluchowski equations \cite{kaniadakis,frank,nfp}. If the particles
interact via weak long-range forces, we obtain generalized mean field
kinetic equations \cite{pre,nfp}. The general study of these
equations, which combine both long-range interactions and generalized
thermodynamics, is very rich and can lead to a wide diversity of phase
transitions and blow-up phenomena. For example, generalized mean field
Fokker-Planck equations appear in the physics of bacterial populations
driven by chemotaxis (Keller-Segel models)
\cite{ks,nfp}.  More generally, long-range interactions and generalized
thermodynamics can have applications in several domains of physics,
astrophysics and biology \cite{cras}.

\section{Appendix: Second variations and local stability equivalence}

The results of Sec. 5 show that the variational problems (\ref{mf1})
and (\ref{projm}) are equivalent for global maximization. In this
Appendix, adapting the procedure developed by Bouchet \cite{baussois}
for the 2D Euler equation, we show that they are also equivalent for
local maximization. A critical point of (\ref{mf1}) is a local maximum of
$S_{\chi}[\rho]$ at fixed mass, energy and normalization iff
\begin{eqnarray}
\delta^{2}F[\delta\rho]=T\int \frac{(\delta\rho)^2}{2\rho}d{\bf r}d{\bf v}d\eta+\frac{1}{2}\int \delta\overline{f}\delta\Phi d{\bf r}d{\bf v}\ge 0,
\label{app1}
\end{eqnarray}
for all perturbations $\delta\rho$ that conserve mass, energy and normalization at first order. On the other hand, a critical point of (\ref{projm}) is a local maximum of $S[\overline{f}]$ at fixed mass and energy iff
\begin{eqnarray}
\delta^{2}F[\overline{f}]=\frac{1}{2}T\int C''(\overline{f}) (\delta \overline{f})^2 d{\bf r}d{\bf v}+\frac{1}{2}\int \delta\overline{f}\delta\Phi d{\bf r}d{\bf v}\ge 0,
\label{app2}
\end{eqnarray}
for all perturbations $\delta\overline{f}$ that conserve mass and
energy at first order.  To make the connection between the second
order variations (\ref{app1}) and (\ref{app2}), the idea is to project
the perturbation $\delta\rho$ on a suitable space (to be defined) and
write $\delta\rho=\delta\rho_{\|}+\delta\rho_{\perp}$ where
$\delta\rho_{\perp}$ is the orthogonal perturbation
\cite{baussois}.  The perturbation $\delta\rho$ must satisfy $\int
\delta\rho d\eta=0$ and $\int \delta\rho\eta
d\eta=\delta\overline{f}$. We {\it impose} the same constraints on
$\delta\rho_{\|}$, i.e. $\int \delta\rho_{\|} d\eta=0$ and $\int
\delta\rho_{\|}\eta d\eta=\delta\overline{f}$ (we shall see that this
leads naturally to the orthogonality of $\delta\rho_{\|}$ and
$\delta\rho_{\perp}$). Then, writing $\delta\rho_{\|}=\delta
\overline{f} g \rho$ where $g$ is a function to be determined, we must
have $\int g\rho d\eta=0$ and $\int g \rho \eta d\eta=1$. We look for
a solution of the form $g=a+b\eta$ and find that $a+b\overline{f}=0$
and $a\overline{f}+b\overline{f^2}=1$ leading to
$a=-\overline{f}/f_{2}$ and $b=1/f_2$ where $f_2$ is the centered
local variance defined in Eq. (\ref{g1}). Therefore, we write
\begin{eqnarray}
\delta\rho=\frac{\delta\overline{f}}{f_2}(\eta-\overline{f})\rho+\delta\rho_{\perp}, 
\label{app3}
\end{eqnarray}
where $\delta\rho_{\perp}$ ensures that all the perturbations are considered.
By construction, we have $\int \delta\rho_{\perp} d\eta=0$ and $\int
\delta\rho_{\perp}\eta d\eta=0$. Therefore, $\delta\rho_{\perp}$ is orthogonal to $\delta\rho_{\|}$ in the sense that
\begin{eqnarray}
\int \frac{\delta\rho_{\|}}{\rho}\delta\rho_{\perp}d\eta\propto \int (\eta-\overline{f})\delta\rho_{\perp}d\eta=0.
\label{app4}
\end{eqnarray}
Then, we readily obtain
\begin{eqnarray}
\int \frac{(\delta\rho)^{2}}{\rho}d\eta=\int \frac{(\delta\rho_{\perp})^{2}}{\rho}d\eta+\int \frac{(\delta\rho_{\|})^{2}}{\rho}d\eta\nonumber\\
=\int \frac{(\delta\rho_{\perp})^{2}}{\rho}d\eta
+\frac{(\delta\overline{f})^2}{f_2}
=\int \frac{(\delta\rho_{\perp})^{2}}{\rho}d\eta+C''(\overline{f})(\delta\overline{f})^2,
\label{app5}
\end{eqnarray}
where we have used identity (\ref{g11}). Finally, the second variations (\ref{app1}) can be rewritten
\begin{eqnarray}
\delta^{2}F[\delta\rho]=T\int \frac{(\delta\rho_{\perp})^{2}}{2\rho} d{\bf r}d{\bf v}d\eta+\delta^{2}F[\delta\overline{f}]. 
\label{app6}
\end{eqnarray}
If $\delta^{2}F[\delta\overline{f}]\ge 0$ for all perturbations
$\delta\overline{f}$ that conserve mass and energy at first order,
then $\delta^{2}F[\delta\rho]\ge 0$ for all perturbations $\delta\rho$
that conserve mass, energy and normalization at first
order. Alternatively, if there exists a perturbation
$\delta\overline{f}_{*}$ such that
$\delta^{2}F[\delta\overline{f}_{*}]< 0$, by taking $\delta\rho_*$ in
the form (\ref{app3}) with $\delta\overline{f}=\delta\overline{f}_{*}$
and $\delta\rho_{\perp}=0$, we get $\delta^{2}F[\delta\rho_{*}]<
0$. We conclude that $\rho({\bf r},{\bf v},\eta)$ is a local maximum
of $S_{\chi}[\rho]$ at fixed $E$, $M$ and normalization iff
$\overline{f}({\bf r},{\bf v})$ is a local maximum of
$S[\overline{f}]$ at fixed $E$, $M$. Thus: (\ref{mf1})
$\Leftrightarrow$ (\ref{projm}) locally.

We now adapt the same procedure to show that the variational problems
(\ref{na2}) and (\ref{dyneq}) are equivalent for local minimization
(we have already shown in Sec. 11 that they are equivalent for global
minimization). A critical point of (\ref{na2}) is a local minimum of $F[{f}]$
at fixed mass iff
\begin{eqnarray}
\delta^{2}F[{f}]=\frac{1}{2}T\int C''({f}) (\delta {f})^2 d{\bf r}d{\bf v}+\frac{1}{2}\int \delta\rho\delta\Phi d{\bf r}\ge 0,
\label{app7}
\end{eqnarray}
for all perturbations $\delta{f}$ that conserve mass at first
order. On the other hand, a critical point of (\ref{dyneq}) is a local
minimum of $F[{\rho}]$ at fixed mass iff
\begin{eqnarray}
\delta^{2}F[{\rho}]=\int \frac{p'(\rho)}{2\rho} (\delta {\rho})^2 d{\bf r}+\frac{1}{2}\int \delta\rho\delta\Phi d{\bf r}\ge 0,
\label{app8}
\end{eqnarray}
for all perturbations $\delta{\rho}$ that conserve mass. We can always write the perturbation in the form
\begin{eqnarray}
\delta f=\frac{\delta\rho}{\int \frac{d{\bf v}}{C''(f)}}\frac{1}{C''(f)}+\delta f_{\perp}, 
\label{app9}
\end{eqnarray}
where $\delta f_{\perp}$ ensures that all the perturbations are considered.
By construction, we have $\int \delta f d{\bf v}=\int \delta f_{\|} d{\bf v}=\delta\rho$ so that $\int
\delta f_{\perp} d{\bf v}=0$. Therefore, $\delta f_{\perp}$ is orthogonal to $\delta f_{\|}$ in the sense that
\begin{eqnarray}
\int C''(f) \delta f_{\|} \delta f_{\perp} d{\bf v} \propto \int \delta f_{\perp} d{\bf v}=0.
\label{app10}
\end{eqnarray}
Then, we readily obtain
\begin{eqnarray}
\int C''(f)(\delta f)^{2}d{\bf v}=\int C''(f) (\delta f_{\perp})^{2} d{\bf v}+\int C''(f)(\delta f_{\|})^{2}d{\bf v}\nonumber\\
=\int C''(f) (\delta f_{\perp})^{2} d{\bf v} 
+\frac{(\delta\rho)^2}{\int \frac{d{\bf v}}{C''(f)}}.
\label{app11}
\end{eqnarray}
Now, a critical point of (\ref{na2}) is of the form $f=F(\beta\epsilon+\alpha)=f(\epsilon)$ with $\epsilon=v^2/2+\Phi({\bf r})$. If we introduce the density $\rho=\int f d{\bf v}$ and the pressure $p=\frac{1}{d}\int f v^2 d{\bf v}$, we find that $\rho=\rho(\Phi)$ and $p=p(\Phi)$. Eliminating $\Phi({\bf r})$ between these relations, we obtain a barotropic equation of state $p=p(\rho)$, the same as in Sec. 11. We easily establish that
\begin{eqnarray}
\nabla p=\frac{1}{d}\int \frac{\partial f}{\partial {\bf r}}v^2 d{\bf v}=\frac{1}{d}\nabla\Phi\int f'(\epsilon) v^2 d{\bf v}\nonumber\\
=\frac{1}{d}\nabla\Phi \int \frac{\partial f}{\partial {\bf v}}\cdot {\bf v} d{\bf v}=-\nabla\Phi\int f d{\bf v}=-\rho\nabla\Phi,
\label{app12}
\end{eqnarray}  
so that the condition $f=f(\epsilon)$ is equivalent to the condition of hydrostatic equilibrium $\nabla p=-\rho\nabla\Phi$. This implies $p'(\Phi)=-\rho(\Phi)$ and 
\begin{eqnarray}
\frac{p'(\rho)}{\rho}=-\frac{1}{\rho'(\Phi)}=-\frac{1}{\int f'(\epsilon)d{\bf v}}=\frac{T}{\int \frac{d{\bf v}}{C''(f)}},
\label{app13}
\end{eqnarray}  
where we have used the identity $C''(f)=-\beta/f'(\epsilon)$ resulting from $C'(f)=-\beta\epsilon-\alpha$. Combining Eqs. (\ref{app11}) and (\ref{app13}) we finally obtain
\begin{eqnarray}
\delta^{2}F[\delta f]=\frac{1}{2}T\int C''(f) (\delta f_{\perp})^{2} d{\bf r}d{\bf v} +\delta^{2}F[\delta \rho].
\label{app14}
\end{eqnarray}
From this equality, we conclude that $f({\bf r},{\bf v})$ is a local minimum of $F[f]$
at fixed mass iff $\rho({\bf r})$ is a local minimum of $F[\rho]$ at fixed
mass. Thus: (\ref{na2})  $\Leftrightarrow$ (\ref{dyneq}) locally.



\bibliographystyle{aipproc}   

\bibliography{sample}

\IfFileExists{\jobname.bbl}{}
 {\typeout{} \typeout{******************************************}
 \typeout{** Please run "bibtex \jobname" to optain} \typeout{** the
 bibliography and then re-run LaTeX} \typeout{** twice to fix the
 references!}  \typeout{******************************************}
 \typeout{} }


\end{document}